\documentclass[sigconf, prologue,table, authorversion]{acmart}

\usepackage[utf8]{inputenc}
\usepackage{multirow}
\usepackage{graphicx}
\usepackage{textcomp}
\usepackage{xspace}
\usepackage{amsthm}
\usepackage{booktabs}
\usepackage{graphics}
\usepackage{mathtools}
\usepackage{hyperref}
\usepackage[font=small,labelfont=bf]{caption}
\usepackage{algorithm}
\usepackage[noend]{algpseudocode}
\usepackage{subfigure}
\usepackage{subcaption}

\newcommand{\score}{\texttt{S}-core}

\newcommand{\w}{\mathbf{w}}

\newcommand{\squishlist}{
 \begin{list}{$\bullet$}
  { \setlength{\itemsep}{0pt}
     \setlength{\parsep}{3pt}
     \setlength{\topsep}{3pt}
     \setlength{\partopsep}{0pt}
     \setlength{\leftmargin}{1.5em}
     \setlength{\labelwidth}{1em}
     \setlength{\labelsep}{0.5em} } }

\newcommand{\squishlisttwo}{
 \begin{list}{$\bullet$}
  { \setlength{\itemsep}{0pt}
    \setlength{\parsep}{0pt}
    \setlength{	opsep}{0pt}
    \setlength{\partopsep}{0pt}
    \setlength{\leftmargin}{2em}
    \setlength{\labelwidth}{1.5em}
    \setlength{\labelsep}{0.5em} } }

\newcommand{\squishend}{
  \end{list}  }
\newcommand{\degg}{\text{deg}}

\newtheoremstyle{sig}
  {}
  {}
  {\itshape}
  {}
  {\scshape}
  {.}
  {.5em}
  {#1 #2\thmnote{\quad(#3)}}

\theoremstyle{sig}

\newtheorem{theorem}{Theorem}
\newtheorem{problem}{Problem}
\newtheorem{corollary}{Corollary}
\newtheorem{dfn}{Definition}

\newtheorem{lemma}[theorem]{Lemma}

\newtheorem{example}{Example}
\newtheorem{property}{Property}[dfn]
\newtheorem{prop}{Proposition}

\definecolor{c1}{HTML}{6E001B} 
\definecolor{c2}{HTML}{6E001B} 
\definecolor{myblue}{HTML}{7BB2DD} 
\definecolor{mygray}{HTML}{DBE2E9}

\usepackage{tikz}
\newcommand*\circled[1]{\tikz[baseline=(char.base)]{
            \node[shape=circle,draw,inner sep=0.3pt, fill=c1] (char) {\textcolor{white}{#1}};}}

\newcommand{\head}[1]{\vspace{1.7mm}\noindent{{\textcolor{c1}{\textbf{#1}}.}}}

\copyrightyear{2024}
\acmYear{2024}
\setcopyright{acmlicensed}\acmConference[KDD '24]{Proceedings of the 30th ACM SIGKDD Conference on Knowledge Discovery and Data Mining}{August 25--29, 2024}{Barcelona, Spain}
\acmBooktitle{Proceedings of the 30th ACM SIGKDD Conference on Knowledge Discovery and Data Mining (KDD '24), August 25--29, 2024, Barcelona, Spain}
\acmDOI{10.1145/3637528.3672011}
\acmISBN{979-8-4007-0490-1/24/08}

\begin{document}

\title{A Unified Core Structure in Multiplex Networks: From Finding the Densest Subgraph to Modeling User Engagement}

\author{Farnoosh Hashemi}
\affiliation{%
    \institution{Cornell University}
  \city{Ithaca}
  \state{NY}
  \country{USA}
  }
\email{sh2574@cornell.edu}

\author{Ali Behrouz}
\orcid{0000-0002-4934-669X}
\affiliation{%
  \institution{Cornell University}
  \city{Ithaca}
  \state{NY}
  \country{USA}
}
\email{ab2947@Cornell.edu}

\renewcommand{\shortauthors}{Hashemi and Behrouz}

\begin{abstract}
  In many complex systems, the interactions between objects span multiple aspects. Multiplex networks are accurate paradigms to model such systems, where each edge is associated with a type. A key graph mining primitive is extracting dense subgraphs, and this has led to interesting notions such as $k$-cores, known as building blocks of complex networks. Despite recent attempts to extend the notion of core to multiplex networks, existing studies suffer from a subset of the following limitations: They \circled{1} force all nodes to exhibit their high degree in the same set of relation types while in multiplex networks some connection types can be noisy for some nodes, \circled{2} either require high computational cost or miss the complex information of multiplex networks, and \circled{3} assume the same importance for all relation types. We introduce \score, a novel and unifying family of dense structures in multiplex networks that uses a function $\texttt{S}(.)$ to summarize the degree vector of each node. We then discuss how one can choose a proper $\texttt{S}(.)$ from the data. To demonstrate the usefulness of \score s, we focus on finding the densest subgraph as well as modeling user engagement in multiplex networks. We present a new density measure in multiplex networks and discuss its advantages over existing density measures. We show that the problem of finding the densest subgraph in multiplex networks is NP-hard and design an efficient approximation algorithm based on \score s. Finally, we present a new mathematical model of user engagement in the presence of different relation types. Our experiments shows the efficiency and effectiveness of our algorithms and supports the proposed mathematical model of user engagement.
\end{abstract}

\begin{CCSXML}
<ccs2012>
   <concept>
       <concept_id>10002950.10003624.10003633</concept_id>
       <concept_desc>Mathematics of computing~Graph theory</concept_desc>
       <concept_significance>500</concept_significance>
       </concept>
   <concept>
       <concept_id>10003752.10003809.10003635</concept_id>
       <concept_desc>Theory of computation~Graph algorithms analysis</concept_desc>
       <concept_significance>500</concept_significance>
       </concept>
 </ccs2012>
\end{CCSXML}



\maketitle
\vspace{-1ex}
\section{Introduction}\label{sec:introduction}
In applications such as biological, social, and financial networks, interactions between objects span multiple aspects. For example, in social networks interactions between people can be social or professional, and  professional interactions can differ according to topics. Accurate modeling of such applications has led to multiplex networks (ML)~\cite{main-ML}, where nodes have interactions in multiple types of connections (a.k.a layers). They have since gained popularity in many applications in social and biological networks, and in opinion dynamics~\citep{Ml-bio, ML-IM, ML-Covid}.

\begin{figure}
    \centering
    \includegraphics[width=0.65\linewidth]{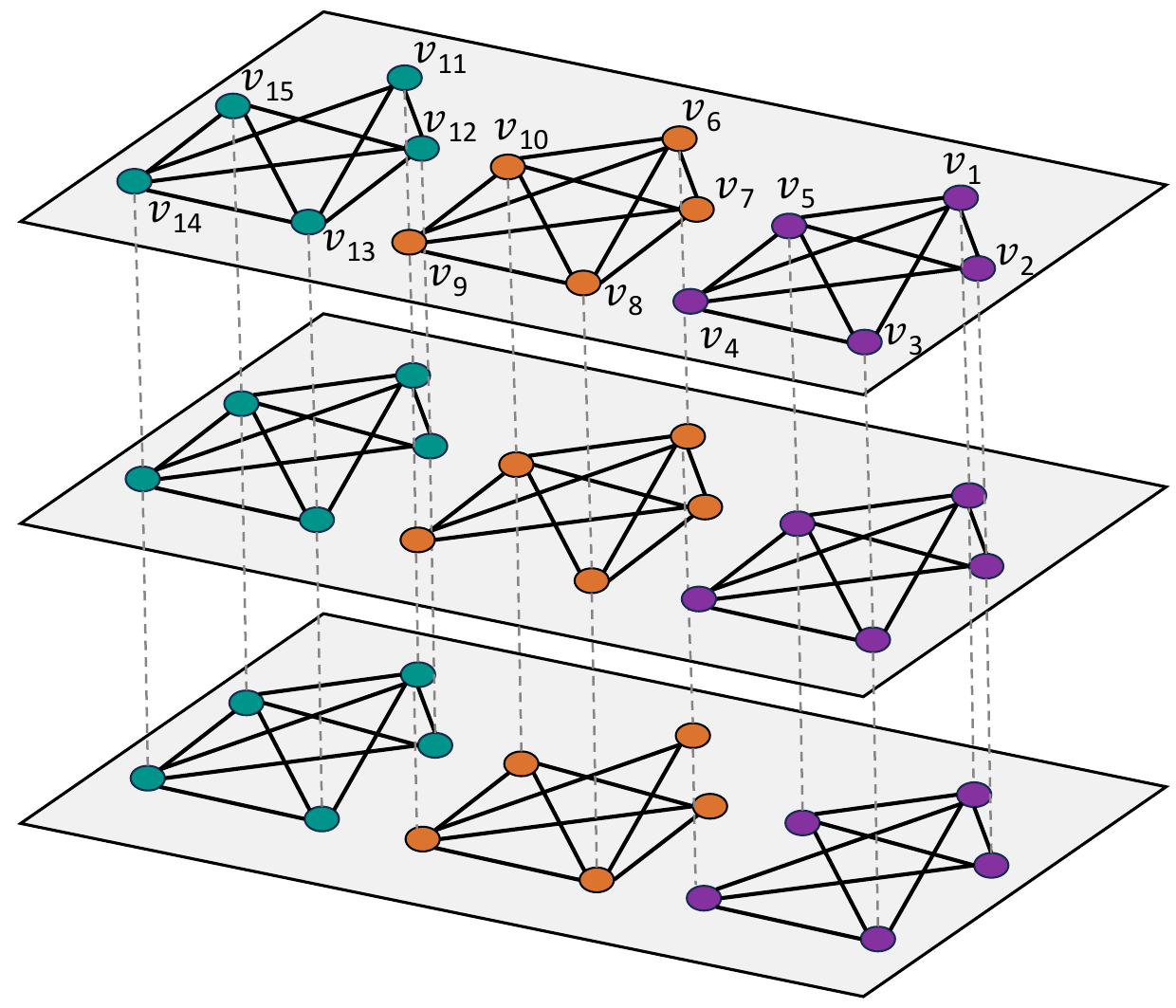}
    \vspace{-2ex}
    \caption{An example of multiplex collaboration network.}
    \label{fig:example}
\end{figure}

\begin{example}
    Figure \ref{fig:example} shows a multiplex collaboration network, where each node is a researcher, each edge is a collaboration, and each layer represents collaborations in an area of research. 
\end{example}

\noindent
Understanding the network topology and finding dense subgraphs are long-standing problems in network science with many applications~\cite{bio-dense, finance-dense, web-dense}. A common method for identifying dense subgraphs is to formulate an objective function (i.e., density) based on \emph{nodes' degree} and solve it via optimization methods~\cite{generalized-MLD, p-mean, densest_first}. While the problem of finding the densest subgraph in simple graphs is a well-studied problem~\cite{densest-survey}, finding dense subgraphs from multiplex networks recently attracts attention~\cite{generalized-MLD, MLcore, FirmCore, FirmTruss}.

\noindent
Despite the recent attempts to extend the known concepts of dense subgraphs to multiplex networks (e.g., \cite{FirmTruss, FirmCore, CoreCube}), existing studies suffer from a subset of the following limitations: \circled{1} They force nodes to satisfy degree constraints in a \emph{fixed subset} of layers, including noisy/insignificant layers~\cite{MLcore}. These layers can be different for each node~\cite{FirmCore} and so this hard constraint can result in missing some dense structures~\cite{FirmCore}. \circled{2} They require exponential running time algorithms, making them infeasible for large real-world graphs. \circled{3} Taking advantage of the complementary information provided by different relation types is challenging as in real-world graphs, different relation types have different importance or roles with respect to the application (see~\cite{flight, admire} for more details). Existing methods treat all relation types equally, causing suboptimal performance and missing information (see \S\ref{sec:experiments}).

\noindent
Besides the above limitations, the complication caused by the definition of degree in multiplex networks makes it challenging to navigate the vast landscape of existing methods and evaluate trade-offs between them in practice. That is, while the degree of each node in simple graphs is represented by a single number, the degree of each node in multiplex networks with $|L|$ layers is an $|L|$-dimensional vector (also called degree vector), in which the $\ell$-th element shows the number of the node's neighbors in layer $\ell$. Accordingly, this high dimensional representation of the degree results in a trade-off between the complexity of the methods and their power to capture the information about the neighborhood of nodes in each layer. For example, \citet{MLcore} suggest using the entire degree vector in objective function, which results in an \emph{exponential time} decomposition algorithm but \citet{FirmCore} suggest sampling the $\lambda$-th largest element of degree vector as its representative, which results in linear time decomposition algorithms. While the former uses all the information provided by the degree vectors, the later methods miss all the information about the nodes' neighbors in different relation types except the $\lambda$-th largest value. Accordingly, it can be challenging to choose a proper objective function to extract dense structures in practice, and in many cases, it requires paying attention to the data and available computational capacity.

\noindent
To mitigate the above limitations, we present \score, a new unifying family of dense subgraphs in multiplex networks. The main intuition of \score s is to use a summarizer function, $\texttt{S}(.)$, and summarize the degree vector of each node into a low dimensional space, mitigating time inefficiency and avoiding too hard constraints. Given a $d$-dimensional threshold vector $\textbf{k}$, we define $(\mathbf{k}, \texttt{S})$-core as a maximal subgraph in which each node $u$ has summrized degree of at least $\mathbf{k}$ within the subgraph. Interestingly, \score s includes existing families of dense multiplex subgraphs as its special cases when choosing different summarizer $\texttt{S}(.)$. We discuss how one can choose a proper $\texttt{S}(.)$ based on the network topology. We propose efficient algorithms to find all possible \score s and show their scalability to graphs with millions of connections (see \S~\ref{sec:experiments}). Finally, we focus on two applications of \score s: \circled{i} Finding the densest multiplex subgraph, and \circled{ii} modeling the user engagement in multiplex networks:

\head{Densest Multiplex Subgraph}
 The main challenge to define the density of a subgraph in multiplex networks is the trade-off between high density and the number of layers in which the high density holds. To this end, existing studies~\cite{MLcore, FirmCore} model the trade-off as a maximin optimization of the average density objective and use a parameter $\beta > 0$ to add a penalty for choosing small subset of layers. While these formulations allow us to control the trade-off, their main drawback is that they force \emph{all} nodes within the subgraph to exhibit their high degree in a \emph{fix} subset of layers. In multiplex networks different layers for different nodes might be noisy/insignificant~\cite{FirmCore, MLcore}. We present a new density measure in multiplex networks, and define the density as the average of maximin optimization of nodes' degree (instead of maximin optimization of the average degree~\cite{FirmCore, MLcore}). This will allow nodes to exhibit their high degree in different and flexible subset of layers. We show that this optimization problem is NP-hard, and use the densest \score{} to approximate the problem of finding the densest subgraph. Not only this approximation algorithm provide effective guarantee for our density formulation, but it also can provide an approximate solution to the problem of densest subgraph with respect to existing density measures~\cite{FirmCore, MLcore} with approximation guarantee that matches the best existing algorithms.

\head{User Engagement}
A fundamental question in understanding social networks is that ``how users decide to engage in a social network?''. Several studies~\cite{anchored} model this as a simultaneous game where each user decides to remain engage or drop out. While these models assume a single type of interaction, in complex social systems, users have different types of interactions and each interaction type has its own effect on the engagement of the users. For example, on Instagram, engagement in sharing posts, stories, and/or sending messages are different for each user. Inspired by \citet{anchored}, we model user engagement in each type of connection as a simultaneous game ($|L|$ games in total), in which each user decides to remain engage or drop out in that specific type of interaction. We show that \score s are unique maximal equilibriums of this game, and provide empirical evidences to support the model design.

\head{Summary of Contributions} 
\circled{1} We present a new family of dense structures in multiplex networks, \score s, that unifies existing degree-based families of dense multiplex subgraphs, using a single function $\texttt{S}(.)$. \circled{2} We provide detailed discussion of how in practice one can choose $\texttt{S}(.)$ to avoid information loss. \circled{3} We introduce a new, efficient, and powerful variant of \score{}, WFirmCore, and show its useful properties. \circled{4} We develop efficient algorithms for the general case (arbitrary function $\texttt{S}(.)$) and WFirmCore. \circled{5} We present a new density objective in multiplex networks that mitigates the hard constraints of existing density measures, leading to finding more cohesive subgraphs. \circled{6} We discuss the hardness of this problem and design an approximation algorithm using the densest \score{} with provable guarantee. We further show that this guarantee is valid when using existing density measures, which matches the best approximation factor for this problem. \circled{7} We further use \score s to model user engagement in multiplex networks as $|L|$ (number of interaction types) simultaneous game in which each user decides to remain engage or drop out in every single type of interaction. We show that \score s are unique maximal equilibriums of this game and provide empirical evidences to support the model design. Proofs, toy examples for all concepts and algorithms, and additional experiments are in Appendix.

\head{Notation}
Let $G = (V, E, L, \w)$ be a multiplex graph, where $V$ is the set of nodes, $L$ is the set of layers (each layer is a graph corresponds to a specific relation type), $E \subseteq V \times V \times L$ is the set of edges, and $\w(.): V \times L \rightarrow \mathbb{R}_{\geq 0}$ is a function that assigns a weight to each layer with respect to a node. The set of neighbors of node $v$ in layer $\ell$ is denoted $N_\ell(v)$ and the degree of $v$ in layer $\ell$ is $\degg_\ell (v) = |N_\ell(v)|$. We use vector $\degg(v)$ to refer to $v$'s degree in each layer (i.e., $(\degg(v))_{\ell} = \degg_\ell (v)$). For a set of nodes $H \subseteq V$,  $\degg^H_{\ell}(v)$ is the degree of $v$ in this subgraph. Given a vector $\textbf{v} = [\textbf{v}_1, \dots, \textbf{v}_d]$, we use Top-$\lambda$ element to refer to $\lambda$-th largest element in $\textbf{v}$. Therefore, we use Top-$\lambda$ degree of $u$ to refer to Top-$\lambda$ element in $u$'s degree vector. The table of notations is in Appendix~\ref{app:notation}.

\section{Related Work and Backgrounds}\label{sec:related-work}
Additional related work and backgrounds is in Appendix~\ref{app:backgrounds} and~\ref{app:rw}.

\head{Dense Subgraph Mining}
 Several variants of the densest subgraph problem with different objective functions have been designed in simple networks~\cite{edge-color, f-density, densest_first, p-mean, p-mean2}. Recently, \citet{p-mean} unifies most existing density objective functions and suggests using $p$-mean of node degrees within the subgraph as its density. In multiplex networks, \citet{densest-common-subgraph} formulate the densest common subgraph problem and develop a linear-programming formulation. \citet{azimi-etal} extended $k$-core to multiplex networks: given an $|L|$-dimensional vector $\mathbf{k} = [k_\ell]_{\ell \in L}$, the ML $\mathbf{k}$-core is a maximal subgraph, in which each node in layer $\ell$ has at least $k_\ell$ neighbors. \citet{gcore} extends this formulation to multilayer networks with inter-layer connections. \citet{MLcore} propose algorithms to find all possible $\mathbf{k}$-cores, and generalized the density measure of \citet{densest-common-subgraph} as follows:

\begin{dfn}[Multilayer Density \cite{MLcore}]\label{dfn:previous-density}
    Given $\beta > 0$, the ML density of subgraph $G[H]$ is defined as:
    \begin{equation}
        \rho(H) = \max_{\hat{L} \subseteq L} \min_{\ell \in \hat{L}} \frac{|E_\ell[H]|}{|H|} |\hat{L}|^\beta.
    \end{equation}
\end{dfn}
\noindent
 \citet{FirmCore} introduce FirmCore as a maximal subgraph in which every node is connected to at least $k$ other nodes within that subgraph, in each of at least $\lambda$ individual layers. Recently, variants of FirmCore, based on triadic closure~\cite{FirmTruss} and $p$-mean~\cite{generalized-MLD}, has been designed to achieve more cohesive structures. Finally, \citet{mlcore-stoc} design an LP-based algorithm to find a stochastic solution for the multiplex densest subgraph problem.

\noindent
All these methods treat all the layers the same, consider too hard degree constraint for nodes in all the layers, or assume pre-defined patterns (which does not necessarily fit all networks in different domain). 
Moreover, \score s are unifying family, meaning that most existing methods are its special cases. They help to navigate the vast landscape of existing methods and evaluate trade-offs between them in practice.

\head{Modeling User Engagement}
Mathematical modeling of user engagement in social networks has attracted attention during the past two decades~\cite{social-payoff, ellison1993learning, blume1993statistical}. \citet{social-payoff} present a payoff structure when the network topology is either complete, a cycle, or a star. Several economic models discuss positive network effects of participation in complete graphs~\cite{katz1985network, arthur1989competing} and competing behavior setting~\cite{ellison1993learning, blume1993statistical}. \citet{anchored} model this problem as a single simultaneous in which each user decides to remain engage or drop out based on its number of active friends. All these studies are different from our model as they have focused on either \circled{i} specific graph topology, and/or \circled{iii} single type of interaction in the network.

\section{\score{} in Multiplex Network}\label{sec:score}
The main intuition of $k$-cores in simple graphs is to decompose the graph into hierarchical structures, in which each node has sufficient number of neighbours. However, the high dimensional degree vectors in multiplex networks makes it challenging to define universally accepted notion of core in multiplex networks. Simply considering the entire degree vector for all nodes \circled{1} can cause computational inefficiency and \circled{2} can be too hard constraint~\cite{FirmCore}. To address these challenges, we present \score{} structures:

% Given a multiplex network $G$, we aim to use a function $\texttt{S}(.)$ that summarizes each degree vector into a low dimensional space. Next, we consider different core numbers for each element of summarized degree vector. This summarization technique can address three main challenges: \circled{i} The summarizing function $\texttt{S}(.)$ is an arbitrary function on degree vectors that might combine node degrees in different layers by considering their dependencies. Accordingly, it can mitigates the limitation of $\mathbf{k}$-core~\cite{MLcore} that sees each relation type separately, missing layer dependencies. \circled{ii} The summarizing function $\texttt{S}(.)$ encode degree vectors to a low dimensional space, making algorithms of finding all core structures faster.
% \circled{iii} It is flexible and can capture more complex dependencies of nodes degrees in different layers than some models like FirmCore~\citep{FirmCore}. That is, as we discuss in \S~\ref{sec:proper-s}, FirmCore \emph{might} miss information about the degree of nodes in each relation type (layer) as it only consider a Top-$\lambda$ degree of each node. However, the summarizing function can summarize nodes degree to $d$-dimensional space, and make sure that the summarized vector carries all the information we need (see \S\ref{sec:proper-s}). We next, formally define \score{} as follows: 

\begin{dfn}[Summarized Core]
    Given a multiplex network $G = (V, E, L, \w)$, a non-decreasing function $\texttt{S}: \mathbb{Z}_{\geq 0}^{|L|} \times \w \rightarrow \mathbb{R}_{\geq 0}^{d}$ that summarizes degree vector of nodes, and a $d$-dimensional vector $\mathbf{k} = [k_{i}]_{1 \leq i\leq d}$, the $(\mathbf{k}, \texttt{S})$-core ($\mathbf{k}$-\score{} for short) of $G$ is a maximal subgraph $H = G[C_{k}] = (C_{k}, E[C_{k}], L)$ such that for each node $v\in C_k$ we have $\texttt{S}(\deg^{C_{\mathbf{k}}}(v), \w)_{i} \geq k_i$ for all $1 \leq i \leq d$. We refer to vector $\mathbf{k}$ as the \texttt{SC} vector index (\texttt{SCV} index for short) of $H = G[C_{k}]$.
\end{dfn}
\vspace{1ex}

% \head{Equivalence Results and Special Cases} In this part, we discuss how choosing different summarizer functions can result in different concepts of cores in multiplex networks. 

\begin{lemma}~\label{lemma:special-case}
    All multilayer $\textbf{k}$-core~\cite{MLcore}, FirmCore~\cite{FirmCore}, and CoreCube~\cite{CoreCube} are special cases of \score s.
\end{lemma}

% \begin{remark}
%     Given a subset of layers $\hat{L} \subseteq L$, let $\w = \mathbf{1}$ and $S(\mathbf{X}) = \mathbf{X}_{\hat{L}}$ (corresponding elements to $\hat{L}$), and $\mathbf{k} = [k, k, \dots, k]_{1 \times |\hat{L}|}$ then $\mathbf{k}$-\score{} is equivalent to CoreCube~\cite{CoreCube}. 
% \end{remark}

% \begin{remark}
%     Given $\lambda \in \mathbb{N}$, let $S(\mathbf{X}) = \text{Top-}\lambda(\mathbf{X})$, then $\mathbf{k}$-\score{} is equivalent to FirmCore~\cite{FirmCore}. 
% \end{remark}

% \head{Properties}
% Next, we discuss the nice properties of \score s: 
% First, note that for each \score{} $G[C]$, there might be several \texttt{SCV} index but there is a unique \texttt{SCV} index that is not dominated by any other \texttt{SCV} indices of $G[C]$: 

\begin{dfn}[Maximal \texttt{SCV} Index]
    Given a multiplex network $G = (V, E, L, \w)$, a function $\texttt{S}: \mathbb{Z}_{\geq 0}^{|L|} \times \w \rightarrow \mathbb{R}_{\geq 0}^{d}$, and $G[C]$ be a \score{} of $G$, let $\mathbf{k} = [k_{i}]_{1 \leq i\leq d}$ be a \texttt{SCV} index of $G[C]$. $\mathbf{k}$ is called maximal \texttt{SCV} index of $G[C]$ if there does not exist any \texttt{SCV} index $\mathbf{k}' = [k'_{i}]_{1 \leq i\leq d}$ of $G[C]$ such that $\forall i \in \{1, \dots, d\}$ we have $k_{i} \geq k'_{i}$ and $\exists j \in \{1, \dots, d\}$ such that $k_{j} > k'_{j}$.
\end{dfn}

\begin{prop}\label{prop:maximal-scv}
    The maximal \texttt{SCV} index for each \score{} exists and is unique.
\end{prop}

\begin{prop}[Uniqueness]\label{prop:unique-core}
    Given a function $\texttt{S}(.)$ and $\mathbf{k} = [k_{i}]_{1 \leq i\leq d}$, the $\mathbf{k}$-\score{} of $G$ is unique.
\end{prop}

\begin{prop}[Hierarchical Structure]\label{prop:hierarchical1}
    Given a multiplex network $G = (V, E, L, \w)$ and two \score s $G[C_{\mathbf{k}}]$ and $G[C_{\mathbf{k}'}]$ with coreness vectors $\mathbf{k} = [k_{i}]_{1 \leq i\leq d} $ and $\mathbf{k}' = [k'_{i}]_{1 \leq i\leq d}$, respectively. If $\forall i \in \{1, \dots, d\}: k_i \geq k'_i$ then $G[C_{\mathbf{k}}] \subseteq G[C_{\mathbf{k}'}]$.
\end{prop}

\noindent
Based on the above hierarchical structure property of \score s, we define skyline \score{} vector indices that can help us to design efficient algorithms for the general case:

\begin{dfn}[Skyline \texttt{SCV} Indices]
    The skyline \texttt{SCV} indices of a multiplex network are all \score s with maximal \texttt{SCV} index $\mathbf{k} = [k_{i}]_{1 \leq i\leq d}$ such that there does not exist any other \score s with maximal \texttt{SCV} index of $\mathbf{k}' = [k'_{i}]_{1 \leq i\leq d}$ where $\forall i \in \{1, \dots, d\} : k'_{i} \geq k_{i}$ and $\exists j$ such that $k'_{j} > k_{j}$.
\end{dfn}

\noindent
Decomposing a multiplex graph to \score s requires finding all $\mathbf{k} = [k_i]_{1 \leq i \leq d}$ corresponding to possible distinct and non-empty $\mathbf{k}$-\score. Based on the above properties, there is a nested property in the search space of all \texttt{SCV} indices and traversing all states in the search space is equivalent to find all possible \score s. To this end, we say a \texttt{SCV} index $\mathbf{k}_{\text{child}}$ is a child of $\mathbf{k}_{\text{parent}}$ if there is $\hat{i} \in \{1, \dots, d\}$ such that for all $i \in \{ 1, \dots, d\}\setminus\{\hat{i}\}$ we have $({\mathbf{k}_{\text{child}}})_{i} = ({\mathbf{k}_{\text{parent}}})_{i}$ and $({\mathbf{k}_{\text{child}}})_{\hat{i}} < ({\mathbf{k}_{\text{parent}}})_{\hat{i}}$. Contrary to the search spaces of other families of dense structures discussed in \cite{MLcore, FirmTruss}, \texttt{SCV} indices are non-negative real numbers and their search space is infinite.  However, not all of these \score s are distinct. Next theorem is the key to design efficient and finite-time algorithms to traverse \emph{distinct} \score s:

\begin{theorem}\label{thm:unique-core}
    Given a multiplex network $G = (V, E, L, \w)$ and its \score{} $G[C]$, the unique maximal \texttt{SCV} index of $G[C]$ is a $d$-dimensional vector $\textsc{Scv}(C) = [\min_{u \in C}\texttt{S}(\deg^{C}(u))_i]_{i = 1}^d$.
\end{theorem}

\subsection{How to Choose $\texttt{S}(.)$?}\label{sec:proper-s}
A natural question about \score s is how to choose proper $\texttt{S}(.)$ to \emph{efficiently} avoid information loss about the neighborhoods of nodes in different layers. 
We need to recall the main intuition behind core structures in networks. Core structures are densely connected subgraphs that nodes are required to satisfy a degree constraint (i.e., each node must have sufficient number of neighbors within the subgraph).
%In multiplex networks, however, each node has different neighborhoods in different layers and so degrees are vectors instead of numbers. Accordingly, not only considering the entire degree vector for all nodes might cause computational overhead, but it also might be unnecessarily as usually there is a correlation between some individual layers. In \score, function $\texttt{S}(.)$ is used to address these challenges but choosing a bad summarizer can cause sever information loss about the neighborhoods of a node in different layers. 
Therefore, we need to choose function $\texttt{S}(.)$ so it can be a good representative for the degree vector of each node. 

\head{Statistical Inference}
Let $\mathcal{L} = \{ \mathcal{L}_1, \dots, \mathcal{L}_t\}$ be a partition of layers such that the degree distribution of all layers in each partition are the same. We let $\mathbb{P}_{\mathcal{D}_i : \theta_i}$ be the degree distribution of layers in partition $\mathcal{L}_i$. Given $i \in \{1, \dots, t\}$, let $\mathcal{L}_i = \{\ell^{(i)}_1, \dots, \ell^{(i)}_{t_i} \}$. We treat $\deg_{\ell_{1}^{(i)}}(u), \dots, \deg_{\ell_{t_i}^{(i)}}(u)$ as $t_i$ samples from the same distribution $\mathbb{P}_{\mathcal{D}_i : \theta_i}$. We now consider two cases:

\vspace{2mm}
\noindent
\circled{1} Layers are independent (e.g., multiplex brain networks~\cite{FirmTruss, admire}, biological networks~\cite{homo}): In this case, given a node $u \in V$, its degrees in layers $\{\ell^{(i)}_1, \dots, \ell^{(i)}_{t_i} \}$ are i.i.d. samples from a distribution $\mathbb{P}_{\mathcal{D}_i : \theta_i}$ with parameter(s) $\theta_i$. Accordingly, given these samples, the minimal sufficient statistics of $\mathbb{P}_{\mathcal{D}_i : \theta_i}$ is the best statistics to preserve information to inference on parameter(s) $\theta_i$, i.e., we do not lose information about the neighborhoods of nodes in layers in $\mathcal{L}_i$. 

\vspace{1mm}
\noindent
\circled{2} Layers are dependent (e.g., social networks~\cite{Twitter_datasets, Higgs, Friendfeed}, collaboration networks~\cite{dblp}):
Accordingly, degrees of a node in different layers are dependent and simple statistical methods cannot be used. We suggest using data summarization methods (e.g., \cite{elhamifar2017subset}), which summarize dependent data with minimal information loss. 

\noindent
We let the output of either \circled{1} or \circled{2} for $\mathcal{L}_i$ be $\texttt{S}_i(.)$. Finally, we aggregate all found summarizer functions to obtain $\texttt{S}(.)$.

\head{Sampling from the Degree Vector}
Another method to efficiently reduce the dimension of degree vectors, is to sample order statistics from the degree vectors. $\lambda$-th order statistic of a vector is its $\lambda$-th smallest value. Accordingly, given $d$ (choose based on computational capacity), and $\lambda_1, \dots, \lambda_d$, one can sample $\lambda_i$-th order statistics to summarize the degree vector. FirmCore~\cite{FirmCore} is a special case of this method, where $d = 1$ and it uses $(|L|-\lambda)$-th order statistic. 

% The main advantage of this approach is that its output is still a vector of integers, which makes the decomposition algorithm faster.

\head{Learn to Find Dense Structures}
 The main drawback of existing dense subgraph mining methods is that they are based on pre-defined patterns or constraints (e.g., $k$-core~\cite{k-core-first}, $k$-truss~\cite{truss}). Real-world networks, however, are complex in nature and a pre-defined pattern cannot fit all networks in different domains. Machine learning models are powerful tools to learn from the data; however, they mostly focus on classification~\citep{behrouz2024graph}, prediction~\cite{admire}, and regression~\cite{jia2020residual} tasks and their usefulness in finding the network building blocks is still unexplored. Our formulation of \score, can bridge the gap and allows using machine learning methods to learn what is the best patterns in a data-driven manner. The main challenge is the lack of objective function. In the information theoretic perspective, since we aim to learn node representation that are good representative for node's degree, we suggest maximizing the mutual information between the actual degree vector and the encoding of nodes. Given a graph neural network $\textsc{Gnn}(.)$~\cite{gcn}, we maximize the following objective: ($\mathcal{I}(.)$ is mutual information)
 \begin{equation}
     \textsc{Loss} := \frac{1}{|V|}\sum_{u \in V} \mathcal{I}(\textsc{Gnn}(u), \deg(u)).
 \end{equation}

\subsection{Temporal Graphs as Multiplex Networks}\label{app:temporal-graph}
Dynamic systems are every where~\citep{behrouz2024chimera}, and temporal networks are powerful paradigms to model the interactions and their dynamics over time in complex dynamic systems. A temporal graph $G = \{G_1, \dots, G_T \}$ is a graph, in which each edge is associated with a timestamp and each $G_i$ shows the graph snapshot at time $i$~\cite{main-temporal-graph}. Recently, representation of temporal graphs as multiplex networks attracts attention, where each layer of the multiplex network represents a snapshot of the temporal graph~\cite{FirmCore, zhu2022discovering}. 

\begin{lemma}\label{lemma:span-core}
    Given $\lambda$, let $\w_\ell = 2^\ell$ for all $\ell \in L$ and $\texttt{S}(\mathbf{X}) = \text{Top-}\lambda(\mathbf{X})$, then all $(k, \Delta)$-Span-cores~\cite{core-temporal} are special cases of $k$-\score. 
\end{lemma}

\noindent
In temporal graphs, the degree vector of each node is a time series data that shows node's degree over time. Accordingly, to find $\texttt{S}(.)$, one can use summarizing methods for time series data (e.g.,~\cite{ogras2006online}).

% \head{Temporal Networks}
% Temporal networks are powerful paradigms to model the interactions and their dynamics over time in complex systems. They can be seen as special cases of multiplex networks, where each layer is a snapshot of the network. We discuss \score s in temporal graphs, and how the existing core models in temporal networks are their special cases in Appendix~\ref{app:temporal-graph}.

%This formulation is good for future studies to bridge graph learning and dense subgraph mining to understand the building blocks of real-world networks. 

\subsection{\score: Algorithms}\label{sec:algorithms}
Given a vector $\mathbf{k} = [k_{i}]_{1 \leq i\leq d}$, Algorithm~\ref{alg:SFC} finds the $\mathbf{k}$-\score, $\mathcal{C}_{\mathbf{k}}$. We start from the entire graph and in each iteration, we remove a node $u$ that does not satisfy \score's condition, i.e., $\texttt{S}(\deg(u))_i < k_i$ for some $i$. We repeat this process until all remaining nodes satisfies \score's conditions. This algorithm iterates at most $|V|$ times as we stop or remove one node in each iteration. Also, in each iteration, we update the degree vector with $\mathcal{O}(|L|)$ and check the conditions with $\mathcal{O}(d)$. Since $d \leq |L|$, the time complexity is $\mathcal{O}(|V| |L|)$.

\head{Toy Example for Algorithm~\ref{alg:SFC}} In the graph shown in Figure~\ref{fig:example},  for the sake of simplicity assume $\w=1$. Since the graph has three layers, the degree of each node is a vector of size three. Assume that the degree summarization function for this graph is given as $S(\begin{bsmallmatrix}
           x_{1} \\
           x_{2} \\
           x_{3}
         \end{bsmallmatrix}) = \begin{bsmallmatrix}
           min (x_{1},x_{2},x_{3}) \\
           max (x_{1},x_{2},x_{3}) 
         \end{bsmallmatrix}$. 
Also assume $ \mathbf{k}= \begin{bsmallmatrix}
           3 \\
           4 \\
         \end{bsmallmatrix} $ and we want to find $\mathbf{k}-$\score, which is a maximal subgraph in which each node has at least a minimum degree of 3 and at least a maximum degree of 4 across the three layers within the subgraph.  In the first iteration,  $H= V$ and we have $deg(v_6) = \begin{bsmallmatrix}
           4 \\
           4 \\
           2\\
         \end{bsmallmatrix}$ and  $S( \begin{bsmallmatrix}
           4 \\
           4 \\
           2\\
         \end{bsmallmatrix}) = \begin{bsmallmatrix}
           2 \\
           4 \\
         \end{bsmallmatrix}$. Since $2 < k_1 = 3$, vertex $v_6$ will be removed from $H$. After removing $v_6$, the degree vectors of its neighbors will be updated, resulting in all its neighbors having a minimum degree of less than 3. Consequently, $v_7$, $v_8$, $v_9$, and $v_{10}$ (all orange vertices) will be removed from $H$.  Additionally, since $deg(v_2) = \begin{bsmallmatrix}
3 \\
3 \\
3
\end{bsmallmatrix}$, $S( \begin{bsmallmatrix}
3 \\
3 \\
3
\end{bsmallmatrix}) = \begin{bsmallmatrix}
3 \\
3 
\end{bsmallmatrix}$, and $3 < k_2 = 4$, vertex $v_2$ will be removed from $H$, and the degree vectors of all its neighbors will be updated. Consequently, $v_1,v_3,v_4$ and $v_5$ (purple vertices) will have a maximum degree of less than 4 and will be discarded from $H$. The nodes remaining in $H$, which are only the green nodes, have a minimum degree of at least 3 and maximum degree of at least 4; therefore, they constitute the $\mathbf{k}$-\score.

\begin{algorithm}[t]
    \small
    \caption{Finding $\mathbf{k}$-\score}
    \label{alg:SFC}
    \begin{algorithmic}[1]
        \Require{A multiplex network $G = (V, E, L, \w)$, a set of nodes $H \subseteq V$, a non-decreasing function $\texttt{S}: \mathbb{Z}_{\geq 0}^{|L|} \times \w \rightarrow \mathbb{R}_{\geq 0}^{d}$, and a vector $\mathbf{k} = [k_{i}]_{1 \leq i\leq d}$.}
        \Ensure{The $\mathbf{k}$-\score{} of $G$.}
        \While{$\exists u \in H$ and $\exists i$: $\: \texttt{S}(\deg^H(u))_i < k_i$}
        \State $H \leftarrow H \setminus \{ u \}$;
        \EndWhile
        % \State $\mathcal{C}_{\mathbf{k}} \leftarrow H$; \\
        \Return $G[H]$;
    \end{algorithmic}
\end{algorithm}

\begin{algorithm}[t]
    \small
    \caption{\score{} Decomposition of Multiplex Networks}
    \label{alg:S-decomposition}
    \begin{algorithmic}[1]
        \Require{A multiplex network $G = (V, E, L, \w)$, and a non-decreasing function $\texttt{S}: \mathbb{Z}_{\geq 0}^{|L|} \times \w \rightarrow \mathbb{R}_{\geq 0}^{d}$.}
        \Ensure{All distinct and non-empty $\mathbf{k}$-\score{} of $G$.}
        \State $\mathbf{Q}_{\textsc{Bfs}} \leftarrow \{ [0]_{d} \}$;
        \State $\mathbf{Q}_{\textsc{Dfs}} \leftarrow \bigcup_{i \in \{1, \dots, d\}} \{ \mathbf{k} | \mathbf{C}_{\mathbf{k}} \in \textsc{Dfs}\text{-}\textsc{Path}\left(G, \texttt{S}, \mathbf{Q}_{\textsc{Bfs}}, i\right)\}$; \Comment{Algorithm~\ref{alg:dfs-path}}
        \State $\mathcal{C}$ keeps visited \score s until now, i.e., $\mathcal{C} \leftarrow \{ C_{\mathbf{k}} | \mathbf{k} \in \mathbf{Q}_{\textsc{Dfs}}\}$;
        \While{$\mathbf{Q}_{\textsc{Bfs}} \setminus \mathbf{Q}_{\textsc{Dfs}} \neq \emptyset$}
            \State Pick and remove $\mathbf{k} = [k_{i}]_{i \in \{1, \dots, d\}}$ from queue $\mathbf{Q}_{\textsc{Bfs}}$;
                    \If{$\mathbf{k} \not \in \mathbf{Q}_{\textsc{Dfs}}$}
                    \State $\mathbf{C}_{\mathbf{k}} \leftarrow$\score$(G, \bigcup_{\tilde{\mathbf{k}} \in \mathcal{P}(\mathbf{k})} \mathbf{C}_{\tilde{\mathbf{k}}}, \mathbf{k})$; \Comment{ Algorithm~\ref{alg:SFC}},~Corollary~\ref{cor:parents}
                    \If{$\mathbf{C}_{\mathbf{k}} \neq \emptyset$}
                        \State $\mathcal{C} \leftarrow \mathcal{C} \cup \{ \mathbf{C}_{\mathbf{k}} \}$;
                        \State $\mathbf{Q}_{\textsc{Dfs}} \hspace{-1mm}\leftarrow \mathbf{Q}_{\textsc{Dfs}}\hspace{-0.5mm} \cup \hspace{-0.5mm}\{ \mathbf{k}, [\underset{u \in \mathbf{C}_{\mathbf{k}}}{\min}\texttt{S}(\deg^{\mathbf{C}_{\mathbf{k}}}(u))_i]_{i = 1}^d \}$; \Comment{Theorem~\ref{thm:unique-core}}
                    \EndIf
                \Else
                    \For{$i = 1, \dots, d$}
                        \State ${k_{i}}^{\text{new}} \leftarrow k_i$; $u^* \leftarrow \emptyset$; $\hat{\mathbf{C}}_{\mathbf{k}} \leftarrow \mathbf{C}_{\mathbf{k}}$;
                        \While{${k_{i}}^{\text{new}} = k_i$} \Comment{Theorem~\ref{thm:unique-core}}
                            \State $\hat{\mathbf{C}_{\mathbf{k}}} \leftarrow \hat{\mathbf{C}_{\mathbf{k}}} \setminus \{u^*\}$
                            \State $u^* \leftarrow \underset{u \in \mathbf{C}_{\mathbf{k}}}{\arg\min}\texttt{S}(\underset{v \in \mathbf{C}_{\mathbf{k}} \setminus \{u\}}{\min}\deg^{\mathbf{C}_{\mathbf{k}} \setminus \{u\}}(v))_i $; 
                            \State $k_i^{\text{new}} \leftarrow \texttt{S}(\underset{v \in \mathbf{C}_{\mathbf{k}} \setminus \{u^*\}}{\min}\deg^{\mathbf{C}_{\mathbf{k}} \setminus \{u^*\}}(v))_i$;
                        \EndWhile
                        \State $\hat{\mathbf{k}} \leftarrow [k_1, \dots, {k_{i}}^{\text{new}}, \dots, k_d]$;
                        \State enqueue $\hat{\mathbf{k}}$ into $\mathbf{Q}_{\textsc{Bfs}}$; $\mathcal{P}(\hat{\mathbf{k}}) \leftarrow \mathcal{P}(\hat{\mathbf{k}}) \cup \{\mathbf{k}\}$; \Comment{Theorem~\ref{thm:unique-core}}
                    \EndFor
                \EndIf
        \EndWhile
        \Return $\mathcal{C}$;
    \end{algorithmic}
\end{algorithm}

\begin{algorithm}[t]
    \small
    \caption{\textsc{Dfs}-\textsc{Path}}
    \label{alg:dfs-path}
    \begin{algorithmic}[1]
        \Require{A multiplex network $G = (V, E, L, \w)$, a set $H \subseteq V$, $\texttt{S}: \mathbb{Z}_{\geq 0}^{|L|} \times \w \rightarrow \mathbb{R}_{\geq 0}^{d}$, a vector $\mathbf{k} = [k_{i}]_{1 \leq i\leq d}$, and an index $i \in \{1, \dots, d\}$.}
        \Ensure{The set of all the \score s of $G$ varying the $i$-th element of $\mathbf{k}$.}
        \State $\mathcal{C}(\mathbf{k}, i) \leftarrow \emptyset$; $\mathbf{B} \leftarrow \emptyset$; $\texttt{Index} \leftarrow \emptyset$
        \For{$u \in H$}
            \State $\mathbf{B}[\texttt{S}(\deg^{H}(u))_{i}] \leftarrow \mathbf{B}[\texttt{S}(\deg^{H}(u))_{i}] \cup \{ u \}$;
            \State $\texttt{Index} \leftarrow \texttt{Index} \cup \{\texttt{S}(\deg^{H}(u))_{i}\}$;
        \EndFor
        \For{$k \in \texttt{Index}$}
            \While{$\mathbf{B}[k] \neq \emptyset$}
                \State remove a node $u$ from $\textbf{B}[k]$; $H \leftarrow H \setminus \{u\}$;
                \For{$(u, v, \ell) \in E[H]$ and $\texttt{S}(\deg^H(v))_i \geq k$}
                    \State $\mathbf{B}[\texttt{S}(\deg^{H}(v) + 1)_{i}] \leftarrow \mathbf{B}[\texttt{S}(\deg^{H}(v) + 1)_{i}] \setminus \{v\}$;
                    \State $\mathbf{B}[\texttt{S}(\deg^{H}(v))_{i}] \leftarrow \mathbf{B}[\texttt{S}(\deg^{H}(v))_{i}] \cup \{v\}$;
                    \State $\texttt{Index} \leftarrow \texttt{Index} \cup \{\texttt{S}(\deg^{H}(v))_{i}\}$;
                \EndFor
                \For{$j \in \{ 1, \dots, d\} \setminus \{i\}$}
                    \For{$(u, v, \ell) \in E[H]$ and $\texttt{S}(\deg^H(v))_j < k_j$}
                        \State $\mathbf{B}[\texttt{S}(\deg^{H}(v))_{i}] \leftarrow \mathbf{B}[\texttt{S}(\deg^{H}(v))_{i}] \setminus \{v\}$;
                    \EndFor
                \EndFor
            \EndWhile
            \State $\mathcal{C}(\mathbf{k}, i) \leftarrow \mathcal{C}(\mathbf{k}, i) \cup \{H\}$;
        \EndFor
        \Return $\mathcal{C}(\mathbf{k}, i)$;
    \end{algorithmic}
\end{algorithm}

\noindent
Simply traversing \texttt{SCV} indices and applying Algorithm~\ref{alg:SFC} to find \score s results in inefficient performance since we start from the entire graph for different \texttt{SCV} indices. Using \score s' nested property (Proposition~\ref{prop:hierarchical1}), we use the following corollary to improve~efficiency:

\begin{corollary}\label{cor:parents}
    Given a $\mathbf{k}$-\score{}, $C$, let $\mathcal{P}(C)$ be the set of all \score s with maximal \texttt{SCV} indices that are the parents of $\mathbf{k}$ in the search space lattice. We have $C \subseteq \bigcap_{\tilde{C} \in \mathcal{P}(C)} \tilde{C}$.
\end{corollary}

\noindent    
While the corollary suggests using breadth-first search (BFS) traverse over the maximal \texttt{SCV} indices, so the union of \score s corresponds to parents can be used in the next level, a simple BFS traverse causes counting each index as many as its number of parents. Further, at each new level of the search we re-start the decomposition algorithm, which is inefficient. To mitigate it, we use a depth-first search traverse of indices, while inspired by BFS, we also consider the union of \score s corresponds to the parents of each state.

\noindent
Algorithm~\ref{alg:S-decomposition} shows the pseudocode of the \score{} decomposition algorithm. $\mathbf{Q}_{\textsc{Bfs}}$ and $\mathbf{Q}_{\textsc{Bfs}}$ are queues that keep track of search in breadth-first and depth-first manners, respectively. We start from the root (i.e., $[0]_{d}$). To look ahead and traverse indices in DFS manner, we first find \score s corresponds to all indices that only their $i$-th element is non-zero. These \score s has nested property and we only need to look at $i$-th element. Algorithm~\ref{alg:dfs-path} shows this procedure. We use bin-sort~\cite{k-core-algorithm} to keep the $i$-th element of $\texttt{S}(\deg(.))$ of nodes sorted throughout the algorithm and can update them in $\mathcal{O}(1)$ time. We recursively remove nodes in bucket $k$ until all remaining nodes satisfy the \score{} conditions. Note that, here, the indices of buckets are real numbers and so we use set $\texttt{Index}$ to store all indices of non-empty buckets.

\noindent
In Algorithm~\ref{alg:S-decomposition}, after finding \score s corresponds to the above paths (line 2), we start from an index in the \textsc{Bfs} queue. If it has not seen in the \textsc{Dfs} traverse, we use Algorithm~\ref{alg:SFC} to find its corresponding \score. Notably, based on Corollary~\ref{cor:parents}, we start from the union of the  \score s corresponds to the current index parents. Lines 8-10 store non-empty \score s and potential root indices for the \textsc{Dfs} traverse. Lines 12-19 find potential \texttt{SCV} indices. Note that, using Theorem~\ref{thm:unique-core}, it only needs to look at indices whose $i$-th element is the minimum $\texttt{S}(\deg(.))_i$ of nodes within the subgraph. Therefore, for next $\texttt{SCV}$ index, we need to change the $i$-th element to second smallest $\texttt{S}(\deg(.))_i$ within the subgraph (changing to the smallest result in the same \score as the previously found \score).

\begin{lemma}
    There is a multiplex network $G = (V, E, L, \w)$ and function $\texttt{S}(.)$ such that Algorithm~\ref{alg:S-decomposition} takes $\mathcal{O}(|V|^{d + 1} |L| + |E| |V|^{d} )$ time.
\end{lemma}

\subsection{\score's variants: Weighted FirmCore}\label{sec:WFC}
There is a trade-off between the diverse solution space and the time complexity of dense subgraph models in multiplex networks. While \score{} provide the most diverse search space and can unify previous core structures in multiplex networks, it might require exponential time algorithm to find all \score s (e.g., when $d \in \mathcal{O}(|L|)$). Next, we discuss a simple but effective (see Sections~\ref{sec:MDS} and \ref{sec:experiments}) variant of \score s and show that it \circled{1} is potentially more efficient than the general case, and \circled{2} has good quality in both theory and practice.
%, and \circled{3} still includes most existing methods and can provide more diverse search space than FirmCore and its existing variants~\cite{FirmCore, FirmTruss, generalized-MLD}.

\begin{dfn}[Layer-Weighted FirmCore]\label{WFirmCore}
Given a multiplex graph $G$, a non-negative \underline{real-value} threshold $\lambda$, and an integer $k \geq 0$, the $(k, \lambda)$-WFirmCore of $G$ is a maximal subgraph $H = G[C_{k}] = (C_{k}, E[C_{k}], L)$ such that for each node $v\in C_k$ there are some layers with cumulative relative importance of at least $\lambda$ (i.e., $ \exists \{\ell_1, ..., \ell_s\} \subseteq L$ with $\sum_{i = 1}^s \w(v, \ell_i) \geq \lambda$) such that $\deg_{\ell_i}^{C_k}(u) \geq k$, for $1\leq i\leq s$.
\end{dfn}

% \noindent
% While the main intuition behind both FirmCore and WFirmCore is to summarize degree vector into a single number (i.e., cumulative relative importance of layers with high density), considering different weights for different layers with respect to each node provides a more generalize search space for core structures.

\noindent
 When refering to $(k, \lambda)$-WFirmCore, we assume that $\lambda$ is maximal, i.e., for at least one vertex $u$ in $(k, \lambda)$-WFirmCore, there is a subset of layers with cumulative relative importance of exactly $\lambda$ in which $u$ has a degree not less than $k$. Since WFirmCores are special cases of \score s, they have its nice properties. However, changing the value of $\lambda$ requires changing function $\texttt{S}(.)$, which its hierarchical property is unclear.

  \begin{property}[Hierarchical Structure]\label{prop:Hierarchical}
    Given a real-value threshold $\lambda$, and an integer $k \geq 0$, the $(k, \lambda + \epsilon)$-WFirmCore of~$G$ is a subgraph of its $(k, \lambda)$-WFirmCore for any $\epsilon \in \mathbb{R}^{+}$.
\end{property}

\begin{theorem}\label{thm:WFirmCore-hardness}
    WFirmCore decomposition, which is finding all WFirmCores in a multiplex network, is NP-hard. When $\w(.)$ takes integer values, the decomposition can be done in Pseudo-polynomial time.
\end{theorem}

\begin{lemma}~\label{lemma:special-case2}
    All multilayer $\textbf{k}$-core~\cite{MLcore}, FirmCore~\cite{FirmCore}, and CoreCube~\cite{CoreCube} are special cases of WFirmCore.
\end{lemma}

\head{Efficient WFirmCore Decomposition}
 While WFirmCore is a special case of \score, using Algorithm~\ref{alg:SFC} is inefficient as we have nested property with respect to not only $k$, but also $\lambda$ (Property~\ref{prop:Hierarchical}).

\noindent
Given $\lambda$, we define the WFirmCore index of a node $u$, Wcore$_\lambda(u)$, as the set of all $k \in \mathbb{N}$, such that $u$ is part of a $(k, \lambda)$-WFirmCore. We further define Top-$\lambda(\deg(u), \w)$ as the maximum value of $k$ that there are some layers $\{\ell_1, \dots, \ell_t\}$ with a cumulative relative weight (with respect to $u$) of at least $\lambda$ (i.e., $\sum_{i = 1}^t \w(v, \ell_i) \geq \lambda$) in which $\deg_\ell^{H}(u) \geq k$. To calculate the Top-$\lambda(\deg(u), \w)$, we can simply sort vector $\deg^{H}(u)$ and check if the cumulative relative weights of layers in which $u$ has a $\deg^{H}_\ell(u) \geq k$ is $\geq \lambda$. This process takes $\mathcal{O}(|L|\log |L|)$ time. It is simple to see that $u$ can be in at most $(k, \lambda)$-WFirmCore, where $k = $Top-$\lambda(\deg(u), \w)$.   Accordingly, Algorithm \ref{alg:WFirmCore} processes the nodes in increasing order of Top$-\lambda(\deg(u), \w)$. It uses a vector $\mathbf{B}$ of lists such that each element $i$ contains nodes with potential Wcore at most $i$. This technique allows us to keep vertices sorted throughout the algorithm and to update each element in $\mathcal{O}(1)$ time. Algorithm~\ref{alg:WFirmCore} iterates over all given $\lambda_i$s and for each, first initializes $\mathbf{B}$ with the minimum of Top$-\lambda_i(\deg(u), \w)$ and $u$'s index for $\lambda_{i-1}$. The reason is due to the nested property with respect to $\lambda$, if $\text{Wcore}_{\lambda}(u) = k$, then $\text{Wcore}_{\lambda + \varepsilon}(u) \leq k$. We process $\mathbf{B}$'s elements in increasing order. If a node $u$ is processed at iteration $k$, its Wcore$_{\lambda_i}$ is assigned to $k$ and removed from the graph. Therefore, we need to update the degree of its neighbors in each layer, which leads to changing the Top$-\lambda_i(\deg(v), \w)$ of its neighbors and changing their bucket accordingly (lines 10-12). Note that the above algorithm finds all $(k, \lambda_i)$-WFirmCores, given $\lambda_i$ for all $\lambda_i \in \Lambda$: at the end of $(k - 1)$-th iteration, each remaining nodes like $u$ has Top$-\lambda_i(\deg(u), \w) \geq k$ as we removed all nodes with Top$-\lambda(\deg(u), \w)$ less than $k$ in the $(k-1)$-th iteration.

\begin{algorithm}[t]
    \small
    \caption{Finding all $(k, \lambda)$-WFirmCores for all $\lambda \in \Lambda$}
    \label{alg:WFirmCore}
    \begin{algorithmic}[1]
        \Require{\hspace{-0.5mm}A multiplex graph $G\hspace{-0.5mm}=\hspace{-0.5mm}(V, E, L, \w)$ and a sorted set $\Lambda = \{\hspace{-0.5mm}\lambda_1, \dots, \lambda_q \hspace{-0.5mm}\}$.}
        \Ensure{WFirmCore index Wcore$_\lambda(v)$ for each $v \in V$ and $\lambda \in \Lambda$.}
        \For{$\lambda_i \in \Lambda$}
            \State $H \leftarrow V$;
            \For{$v \in H$}
                \State $\texttt{Index}[v] \leftarrow \min\{\text{Top-$\lambda_i$}(\deg(u), \w), {\text{Wcore}_{\lambda_{i - 1}}}(u)\}$;
                \State $\mathbf{B}[\texttt{Index}[v]] \leftarrow B[\texttt{Index}[v]] \cup \{v\}$;
            \EndFor
            \For{$k = 1, 2, \dots, |H|$}
                \While{$\mathbf{B}[k] \neq \emptyset$}
                    \State pick and remove $v$ from $\mathbf{B}[k]$; Wcore$_\lambda(v) \leftarrow k$;
                    \For{$(v, u, \ell) \in E[H]$ and $\texttt{Index}[u] > k$}
                        \State \hspace{-0.5mm}update $\text{Top-$\lambda$}(\deg(u))$; remove $u$ from $\mathbf{B}[\texttt{Index}[u]]$;
                        \State \hspace{-0.5mm}update $\texttt{Index}[u]$; $\mathbf{B}[\texttt{Index}[u]] \hspace{-0.5mm}\leftarrow\hspace{-0.5mm} \mathbf{B}[\texttt{Index}[u]] \cup \{u\}$;
                    \EndFor
                    \State $H \leftarrow H \setminus \{v\}$;
                \EndWhile
            \EndFor
        \EndFor
        \Return WCore;
    \end{algorithmic}
\end{algorithm}

\section{The Multiplex Densest Subgraph}\label{sec:MDS}
As discussed in Section~\ref{sec:related-work}, existing density measures in multiplex networks, assume \circled{1} all layers are complete and important, \circled{2} noisy/insignificant/unimportant layers are the same for all nodes while in many applications, like financial or transportation networks, the importance of each relation type for each node is different~\cite{anomuly}, and \circled{3} all nodes are forced to exhibit their high-density in a fixed subset of layers. To address the limitations, we present a new multiplex density measure that allows layers to have different weights with respect to each node. Moreover, it does not force the densest subgraph to exihibits high degree for all nodes in the same set of layers. The new formulation let each node in the subgraph to exihibits its high degree in different set of layers.

\begin{problem}[Multiplex Densest Subgraph]\label{Problem:density}
    Given a multiplex graph $G=(V,E,L, \w)$, a $\beta > 0$, and function $\rho : 2^V \rightarrow \mathbb{R}^+$:
    \begin{equation} \label{eq:density}
    \rho(S) = \frac{1}{|S|} \sum_{u \in S} \max_{\hat{L} \subseteq L} \min_{\ell \in \hat{L}} \deg_\ell(u) \times \w(u, \ell)  \times \left(\sum_{\ell' \in \hat{L}}\w(\ell')\right)^\beta,
    \end{equation}
    find a subset of vertices $S^* \subseteq V$ that maximizes $\rho$ function, i.e., 
    \begin{equation} \label{eq2}
    S^* = \arg \max_{S \subseteq V} \rho(S).
    \end{equation}
\end{problem}

\noindent
Note that we can simplify the definition of function $\rho(.)$ for a subgraph $S$ as follows:
\begin{align}
    \rho(S) = \frac{1}{|S|} \sum_{u \in S} \max_{\lambda \in \boldsymbol{\Phi}} \text{Top-}\lambda(\deg(u), \w) \lambda^\beta,
\end{align}
where $\boldsymbol{\Phi}$ is the set of weight summation of all subset of layers.

\begin{algorithm}[t]
    \small
    \caption{WFC-Approx}
    \label{alg:Approx1}
    \begin{algorithmic}[1]
        \Require{A multiplex graph $G = (V, E, L, \w)$, and parameter $\alpha \hspace{-0.2mm}\in\hspace{-0.2mm} \{1, \dots, |L| \}$}
        \Ensure{Approximation solution to the densest subgraph problem.}
        \State $\Lambda \leftarrow $ summations of all subsets of layer weights with size $1 \leq s \leq \alpha$;
        \For{$\lambda \in \Lambda $}
        \State $\mathcal{Q}_{\lambda} \leftarrow $ find all $(k, \lambda)$-WFirmCore \Comment{Using Algorithm~\ref{alg:WFirmCore}}
        \State $\hat{C}_{\lambda} \leftarrow$ find the densest $(k, \lambda)$-WFirmCore $\in \mathcal{Q}_{\lambda}$.
        \EndFor
        \Return the densest subgraph among all $\hat{C}_{\lambda}$ for $\lambda \in \Lambda $.
    \end{algorithmic}
\end{algorithm}

\noindent
We followed \citet{MLcore}, and consider a penalty for choosing small number of layers. In fact, term $\left(\sum_{\ell' \in \hat{L}}\w(\ell')\right)^\beta$ encourages the density measure to choose more layers.

\begin{theorem}\label{thm:MDS-hardness}
    The multiplex densest subgraph problem is NP-hard, and cannot be approximated within a constant factor, unless $\texttt{P} = \texttt{NP}$.
\end{theorem}

\head{Approximation Algorithm}
To overcome the complexity of the problem, next, we provide a fast approximation algorithm with provable guarantee. Algorithm~\ref{alg:Approx1} shows the pseudocode of the algorithm. Given a threshold $\alpha$, we first construct a candidate set for the value of $\lambda$. To this end, we consider the set of summations of all possible subsets of layer weights with size $1 \leq s \leq \alpha$, denoted as $\Lambda$. Next, we use Algorithm~\ref{alg:WFirmCore} for each $\lambda \in \Lambda$, and then report the densest WFirmCore as the approximate solution. In our experiments, we observe that $\alpha = \min\{|L|, 10\}$ results in a good approximate solution. We let $\Omega$ be the maximum summation of layer weights over a node, and $S_{\text{SL}}$ be the densest single-layer subgraph among all layers with a minimum degree of $\mu^*$. Let $\psi$ be the maximum value that $(\mu^*, \psi)$-WFirmCore is non-empty:

\begin{theorem}\label{thm:approximation-factor}
    Given $\alpha$, Algorithm~\ref{alg:Approx1} provides $\frac{\min\{\alpha, \psi\}^\beta}{2\Omega^{\beta}}$-approximation solution to the problem of Multiplex Densest Subgraph. 
\end{theorem}

\begin{corollary}[$\alpha$-independent factor]\label{cor:approximation-factor}
    Independent of $\alpha$, when $\w(.) = 1$, Algorithm~\ref{alg:Approx1} provides $\frac{1}{2|L|^\beta}$-approximation solution. 
\end{corollary}

\begin{prop}[Comparison of Density Measures]\label{prop:density-comparison}
    Let $\rho^*(.)$ be the density proposed by \citet{MLcore} (Definition~\ref{dfn:previous-density}) and $\rho(.)$ be our density then for any subgraph $G[H] \subseteq G$ we have:
    \begin{equation}
        \rho(G[H]) \geq \rho^*(G[H]).
    \end{equation}
\end{prop}

\noindent
This result shows the power of our density measure compare to \cite{MLcore}, as it finds any subgraph found by \cite{MLcore} (not vice versa). Also, it shows our algorithm provides $\frac{1}{2|L|^\beta}$-approx solution to the problem of densest subgraph with respect to ML density measure~\cite{MLcore}, which matches its best approximation guarantee \cite{FirmCore, MLcore}.

\begin{table*}[t!]
	\caption{
    The density of the densest different variants of \score s with respect to edge density, ML density~\cite{MLcore}, and our density. The best (resp. the second best) result is highlighted in blue (resp. gray). OOT: Time exceeds 24 hours, OOM: Memory exeeds 100 GB. 
	}
 \vspace{-2ex}
	\label{tab:results}
	\centering
	\scalebox{0.67}{
 % \hspace{-6ex}
 \begin{tabular}{ l l   |  l  | l l     l l l  l l l l l}
			\toprule
		&   &\textbf{Dataset}&\textbf{Homo} & \textbf{Sacchcere} & \textbf{FAO  ~~~~ } & \textbf{Brain ~~~~}& \textbf{DBLP} & \textbf{Amazon} & \textbf{FFTwitter} & \textbf{Friendfeed} & \textbf{StackO} & \textbf{Google+}  \\
		&   & $|V|$ & 18k & 6.5k & 214 & 190 & 513k & 410k & 155k & 510k & 2.6M & 28.9M\\
		&   & $|E|$ & 153k & 247k & 319K & 934K & 1.0M & 8.1M & 13M & 18M & 47.9M & 1.19B\\
            & \textbf{Metric} & $|L|$ & 7 & 7 & 364 & 520 & 10 & 4 & 2 & 3 & 24 & 4 \\
			\midrule
            \midrule
			&   & WFirmCore & 0.58& \cellcolor{mygray}0.46& \cellcolor{myblue}\textbf{0.47} & \cellcolor{myblue}\textbf{0.90} & 0.39& 0.51& 0.59 & \cellcolor{myblue}\textbf{0.48} & \cellcolor{myblue}\textbf{0.53} & \cellcolor{myblue}\textbf{0.84}\\
			& & $\texttt{S}(.) = [\textsc{Max}(.), \textsc{Min}(.)]$ & 0.43 & 0.41 & 0.30 & 0.72 & 0.36 & 0.33 & 0.48 & 0.31 & 0.42 & \underline{OOT}\\
			&  Edge Density $\uparrow$& FirmCore~\cite{FirmCore} & 0.47 & 0.42 & 0.35 & 0.78 & 0.41 & 0.42 & 0.52 & 0.36  & 0.45  & 0.52 \\
			& \multirow{3}{*}{($\frac{\sum_{\ell \in L}\w_\ell|E_\ell[S]|}{\w^* \times {\binom{|S|}{2}}}$)}& ML $\mathbf{k}$-core~\cite{MLcore} & 0.44 & 0.43& \underline{OOM}& \underline{OOM} & 0.37 & 0.40& 0.56 & 0.35 & \underline{OOM} & \underline{OOT}\\
   			& & $\texttt{S}(.) = \textsc{Sum}(.)$ & 0.40 & 0.39& 0.29& 0.70 & 0.32 & 0.37& 0.46& 0.31 & 0.41 & 0.44\\
            & & $\texttt{S}(.) = \textsc{WSum}(.)$ & 0.42 & 0.40& 0.31& 0.73 & 0.35 & 0.34 & 0.49 & 0.32 & 0.40 & 0.47\\
            & & $\texttt{S}(.) = \textsc{MGcn}(.)$~\cite{anomuly} & \cellcolor{mygray}0.69 & \cellcolor{myblue}\textbf{0.66}& 0.38& \underline{OOM} & 0.38& 0.43& 0.52& 0.38 & \underline{OOM} & \underline{OOM}\\
            & & $\texttt{S}(.) = \textsc{MGat}(.)$~\cite{anomuly} & \cellcolor{myblue}\textbf{0.71} & \cellcolor{myblue}\textbf{0.66}& 0.40 & \underline{OOM} & 0.37 & 0.45& 0.51& 0.41 & \underline{OOM} & \underline{OOM}\\
            & & $\texttt{S}(.) = \textsc{Stat}(.)\:$ \textcolor{c1}{[\S~\ref{sec:proper-s}]} & 0.42 & 0.41 & \underline{OOM}& 0.74 & 0.36 & 0.44& 0.47& 0.37 & \underline{OOM} & 0.47\\
			\cmidrule{2-13}
            \cmidrule{2-13}
			&   & WFirmCore & \cellcolor{mygray}31.14 & \cellcolor{myblue}\textbf{28.59} & 1854.07 & 7935.29&  82.91 & 61.38 & 99.26& 216.74 & \cellcolor{myblue}\textbf{118.33} & \cellcolor{myblue}\textbf{173.81}\\
			& & $\texttt{S}(.) = [\textsc{Max}(.), \textsc{Min}(.)]$ & 26.07 & 24.88 & 1469.31& 6932.78 & 74.72 & 38.85 & 96.53& 160.02 & 105.28 & \underline{OOT}\\
			& ML Density $\uparrow$ & FirmCore~\cite{FirmCore} & 29.74 & 25.87 & 1673.18 & 7163.89 & 78.91 & 43.52 & 100.24 & 170.87  & 107.09  & 164.81 \\
			& (\citet{MLcore})& ML $\mathbf{k}$-core~\cite{MLcore} & 27.84 & 26.92& \underline{OOM}& \underline{OOM} & 75.19 & 40.54& 102.37& 164.81 & \underline{OOM} & \underline{OOT}\\
   			& & $\texttt{S}(.) = \textsc{Sum}(.)$ & 23.58 & 22.91& 1419.43& 6846.21 & 71.58 & 37.68& 93.44 & 159.18 & 101.36 & 157.92\\
            & & $\texttt{S}(.) = \textsc{WSum}(.)$ & 25.85 & 24.07 & 1492. 97& 6984.49 & 74.08 & 39.46 & 97.81 & 161.25  & 104.67 & 160.73\\
            & & $\texttt{S}(.) = \textsc{MGcn}(.)$~\cite{anomuly} & \cellcolor{myblue}\textbf{31.50} & \cellcolor{mygray}28.51& 1649.17& \underline{OOM} & 76.65 & 54.52 & 96.48& 184.35 & \underline{OOM} & \underline{OOM}\\
            & & $\texttt{S}(.) = \textsc{MGat}(.)$~\cite{anomuly} & 31.09 & \cellcolor{mygray}28.51& 1752.28& \underline{OOM} & 77.31 & 52.83 & 98.95& 191.28 & \underline{OOM} & \underline{OOM}\\
            & & $\texttt{S}(.) = \textsc{Stat}(.)\:$ \textcolor{c1}{[\S~\ref{sec:proper-s}]} & 25.97& 25.15& \underline{OOM}& 7072.63& 75.07 & 55.38 & 97.24& 91.08 & \underline{OOM} & 162.68\\
			\cmidrule{2-13}
            \cmidrule{2-13}
            &   & WFirmCore & 70.17 & \cellcolor{mygray}58.17& \cellcolor{myblue}\textbf{3044.85}& \cellcolor{myblue}\textbf{10935.29} & \cellcolor{myblue}\textbf{94.36} & \cellcolor{myblue}\textbf{61.38}& \cellcolor{mygray}104.70& \cellcolor{myblue}\textbf{228.09} & \cellcolor{myblue}\textbf{205.98} & \cellcolor{myblue}\textbf{199.62}\\
			& & $\texttt{S}(.) = [\textsc{Max}(.), \textsc{Min}(.)]$ & 60.86 & 50.71 & 2480.63 & 8115.23 & 72.69 & 46.85 & 92.48 & 209.72 & 191.03 & \underline{OOT}\\
			& Our Density $\uparrow$& FirmCore~\cite{FirmCore} & 68.79 & 54.20 & 2718.91& 9017.69 & 82.56 & 57.38 & 101.55 & 220.43 & 198.71 & 180.62 \\
			& (Problem~\ref{Problem:density})& ML $\mathbf{k}$-core~\cite{MLcore} & 63.79 & 51.67& \underline{OOM}& \underline{OOM} & 75.09 & 51.33& 99.71& 219.84 & \underline{OOM} & \underline{OOT}\\
   			& & $\texttt{S}(.) = \textsc{Sum}(.)$ & 60.28 & 47.96 & 2472.58 & 8025.94 & 70.73 & 45.91 & 91.55& 204.57 & 191.22 & 173.38\\
            & & $\texttt{S}(.) = \textsc{WSum}(.)$ & 61.58 & 49.40 & 2499.74 & 8252.33 & 71.18 & 47.38 & 93.58 & 207.49 & 193.14 & 175.26\\
            & & $\texttt{S}(.) = \textsc{MGcn}(.)$~\cite{anomuly} & \cellcolor{mygray}71.43 & \cellcolor{myblue}\textbf{58.33}& 2635.17& \underline{OOM} & 78.68 & 50.04& 95.47& 214.72 & \underline{OOM} & \underline{OOM}\\
            & & $\texttt{S}(.) = \textsc{MGat}(.)$~\cite{anomuly} & \cellcolor{myblue}\textbf{72.39} & \cellcolor{myblue}\textbf{58.33}& 2881.32& \underline{OOM} & 81.25 & 48.87& 97.29& 212.49 & \underline{OOM} & \underline{OOM}\\
            & & $\texttt{S}(.) = \textsc{Stat}(.)\:$ \textcolor{c1}{[\S~\ref{sec:proper-s}]} & 62.51 & 52.20 & \underline{OOM}& 8368.29 & 72.69 & 50.95& 91.68& 211.31 & \underline{OOM} & 176.83\\
            \cmidrule{2-13}
            \cmidrule{2-13}
            &   & WFirmCore &36 &  82& 4219 & 7205 & 872 & 954& 917 & 4788 & 20811 & 74893\\
			& & $\texttt{S}(.) = [\textsc{Max}(.), \textsc{Min}(.)]$ & 31 & 758& 14355& 7584 & 729 & 976& 2980& 6053 & 40172 & \underline{OOT}\\
			& Running Time (s) $\downarrow$& FirmCore~\cite{FirmCore} & \cellcolor{mygray}20 & 41 & 2454 & 3273 & 362 & 394 & 359 & 891  & 8053  & 36027 \\
			& (Decomposition Alg.) & ML $\mathbf{k}$-core~\cite{MLcore} & 57 & 3129& \underline{OOM}& \underline{OOM} & 1283 & 6852& 3082& 14159 & \underline{OOM} & \underline{OOT}\\
   			& & $\texttt{S}(.) = \textsc{Sum}(.)$ & \cellcolor{myblue}\textbf{11} & \cellcolor{myblue}\textbf{17}& \cellcolor{mygray}25& \cellcolor{myblue}\textbf{54} & \cellcolor{myblue}\textbf{62} & \cellcolor{myblue}\textbf{210}& \cellcolor{myblue}\textbf{219}& \cellcolor{mygray}547 & \cellcolor{mygray}628 & \cellcolor{myblue}\textbf{18518}\\
            & & $\texttt{S}(.) = \textsc{WSum}(.)$ & \cellcolor{mygray}24 & \cellcolor{mygray}21& \cellcolor{myblue}\textbf{23}& \cellcolor{mygray}59 & \cellcolor{mygray}68 & \cellcolor{mygray}296& \cellcolor{mygray}307& \cellcolor{myblue}\textbf{526} & \cellcolor{myblue}\textbf{579} & \cellcolor{mygray}19577\\
            & & $\texttt{S}(.) = \textsc{MGcn}(.)$~\cite{anomuly} & 52 & 2062& 30217& \underline{OOM} & 987 & 4570& 1282& 8911 & \underline{OOM} & \underline{OOM}\\
            & & $\texttt{S}(.) = \textsc{MGat}(.)$~\cite{anomuly} & 50 & 2208& 24195& \underline{OOM} & 1199 & 1604& 2735& 5106 & \underline{OOM} & \underline{OOM}\\
            & & $\texttt{S}(.) = \textsc{Stat}(.)\:$ \textcolor{c1}{[\S~\ref{sec:proper-s}]} & 27 & 968& \underline{OOM}& 433 & 775 & 4739& 212& 7946 & \underline{OOM} & 19904\\
			% \cmidrule{2-13}
   %          \cmidrule{2-13}
			%  &   & WFirmCore & 0 & 0& 0& 0 & 0 & 0& 0& 0 & 0 & 0\\
			% & Running& $\texttt{S}(.) = [\textsc{Max}(.), \textsc{Min}(.)]$ & 0 & 0& 0& 0 & 0 & 0& 0& 0 & 0 & 0\\
			% &   Time (s)& FirmCore~\cite{FirmCore} & 0 & 0& 0& 0 & 0 & 0& 0& 0 & 0 & 0 \\
			% & & ML $\mathbf{k}$-core~\cite{MLcore} & 0 & 0& 0& 0 & 0 & 0& 0& 0 & 0 & 0\\
   % 			& & $\texttt{S}(.) = \textsc{Sum}(.)$ & 0 & 0& 0& 0 & 0 & 0& 0& 0 & 0 & 0\\
   %          & & $\texttt{S}(.) = \textsc{WSum}(.)$ & 0 & 0& 0& 0 & 0 & 0& 0& 0 & 0 & 0\\
   %          & & $\texttt{S}(.) = \textsc{Gnn}(.)$ & 0 & 0& 0& 0 & 0 & 0& 0& 0 & 0 & 0\\
   %          & & $\texttt{S}(.) = \textsc{GAT}(.)$ & 0 & 0& 0& 0 & 0 & 0& 0& 0 & 0 & 0\\
			\bottomrule
	\end{tabular}
 }
\end{table*}

\section{A User Engagement Model}\label{sec:user-engagement}
Several studies~\cite{anchored} have modeled user engagement as a simultaneous game where each user decides to remain engage or drop out. However, the main drawback of this approach is that it assumes there is only one type of connection in the network. In complex social systems, users have different type of interactions and each interaction type has its own effect on the engagement of the user. For example, on Instagram, engagement in sharing posts, stories, and/or sending messages are different for each user.
The behavior of each user's friends can affect their type of engagement (whether share a story, post a content, or both). Inspired by \citet{anchored}, we model user engagement in each type of connection as a simultaneous game, in which each user decides to remain engage or drop out. For each user in relation type $\ell$, we define its utility as $\mathbf{u}_{\ell}(v) = |N^+_\ell(v)| - k_\ell$ if it remains engage, and $\mathbf{u}_{\ell}(v) = 0$, if it drops out. Here, $|N^+_\ell(v)|$ is the set of $v$'s neighbour in layer $\ell$ that remains engage. Next, we define the final utility of the user $v$ as $\mathbf{u}^f_{\ell}(v) = \texttt{S}(\mathbf{u}_{\ell_1}(v), \dots, \mathbf{u}_{\ell_{|L|}}(v))$, where $\texttt{S}(.)$ is some function.   

\begin{theorem}\label{thm:equilibrium}
    $\mathbf{k}$-\score{} is the unique maximal equilibrium to the above game, where $(\mathbf{k})_\ell = k_\ell$.
\end{theorem}

\noindent
Having presented the basic theoretical model, next we propose a measure for characterizing the engagement of users:

\begin{dfn}[Node's Engagement]\label{dfn:engagement}
    The engagement level of a node is defined as the maximum L1-norm of its maximal \texttt{SCV} indices:
    \begin{equation}
        \tau(u) = \max_{\mathbf{k} \in \texttt{SCV}(u)} ||\mathbf{k}||_1.
    \end{equation}
\end{dfn}

\noindent
To show the significance of this model, we empirically will answer the following questions in our experimental evaluation: \circled{1} Do we really need to consider different types of interactions? (See \autoref{fig:singlelayer}), \circled{2} Is degree enough to model user engagement? (See Figure~\ref{fig:user-engagement} (Right)), and \circled{3} How does the proposed node's engagement measure work in real networks? (See Figure~\ref{fig:user-engagement} (Left))

\section{Experiments}\label{sec:experiments}
\begin{figure}
    \includegraphics[width=0.22\textwidth]{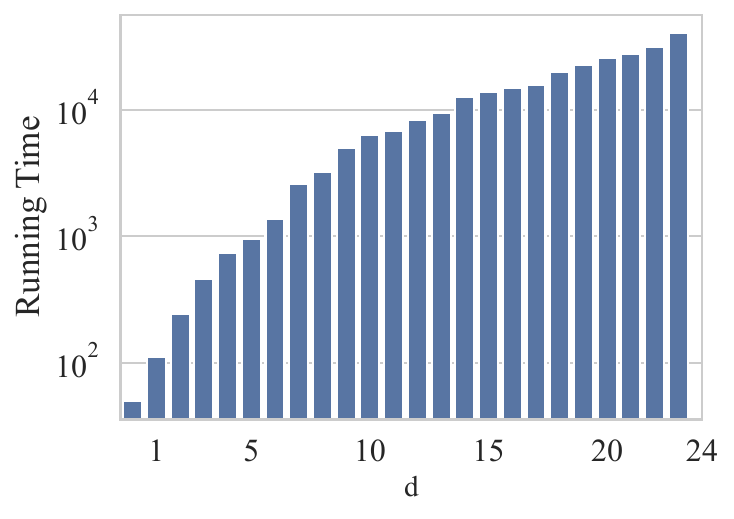}~
    \includegraphics[width=0.23\textwidth]{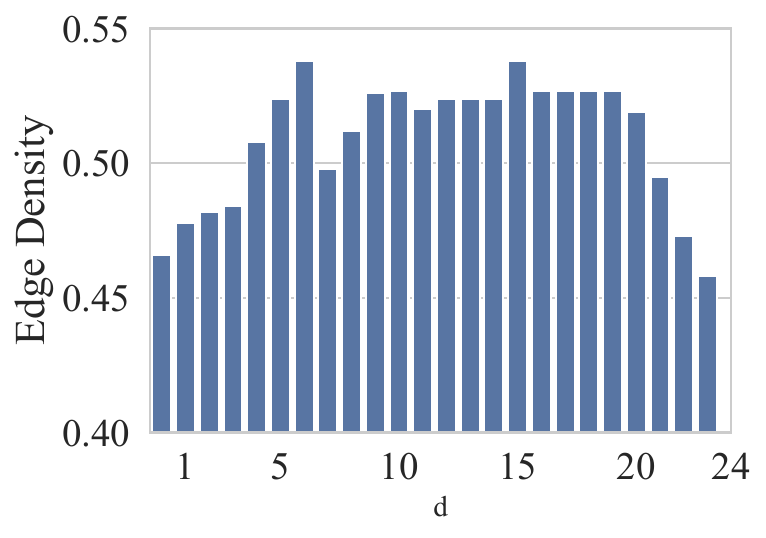}
    \vspace{-3ex}
    \caption{The effect of $d$ on Time (Left), and Density (Right).}
    \label{fig:effect-d}
    \vspace{-2ex}
\end{figure}

\begin{figure}
    \includegraphics[width=0.21\textwidth]{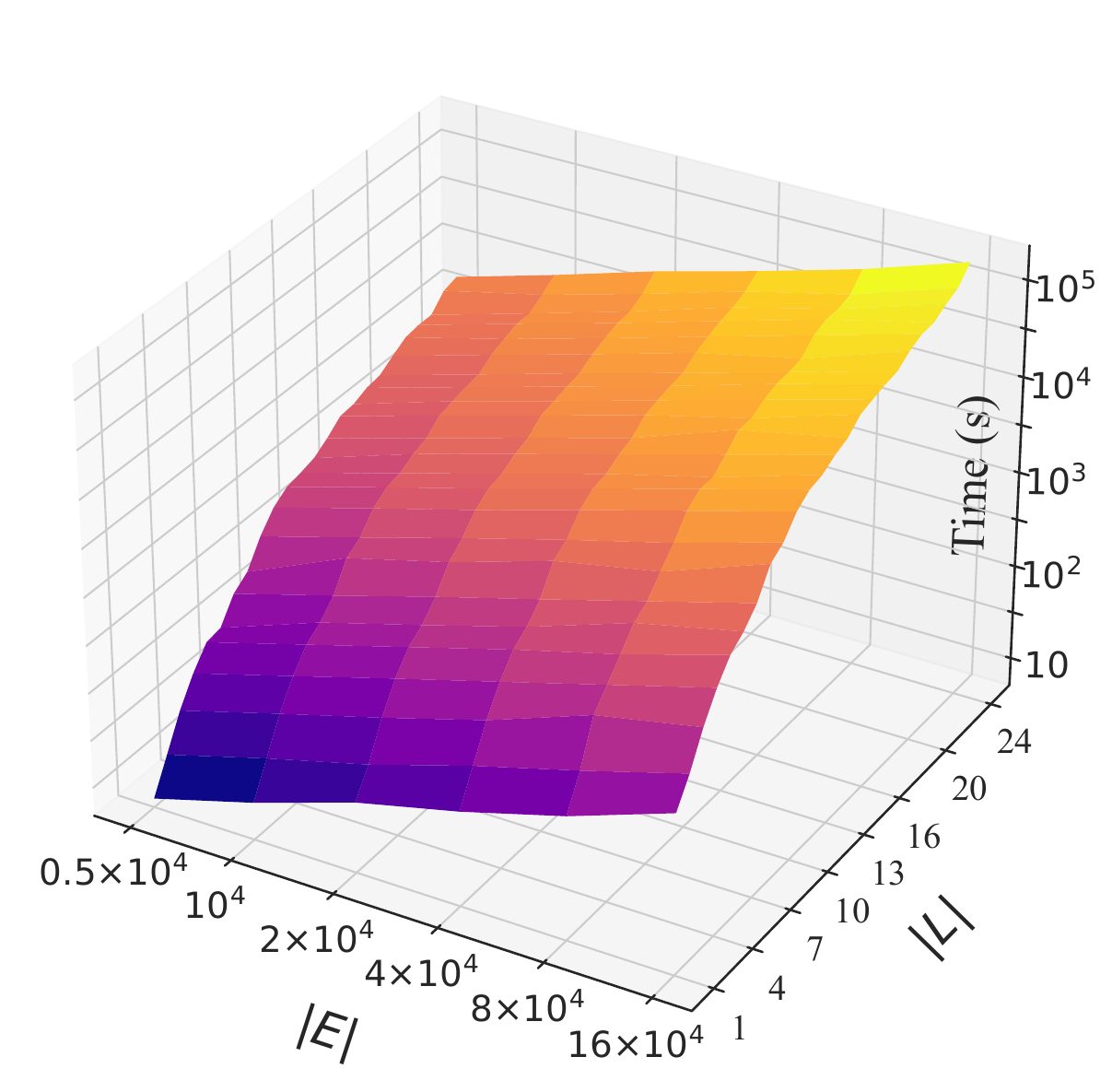}~
    \includegraphics[width=0.205\textwidth]{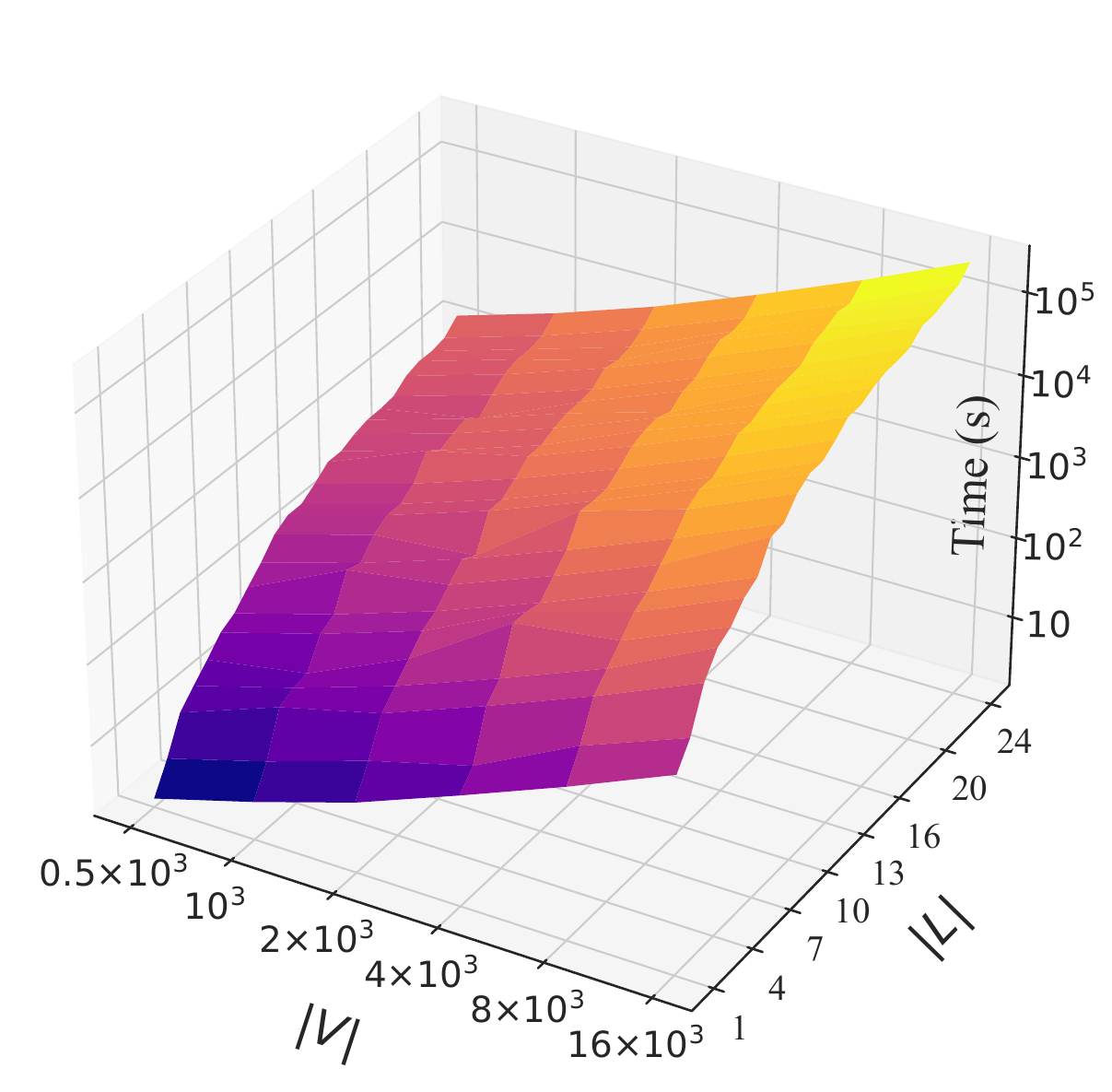}
    % \vspace{-3ex}
    \caption{The effect of $|E|$ (Left), and $|V|$ (Right) on running time.}
    \label{fig:scale}
    \vspace{-2ex}
\end{figure}

In this section, we evaluate our models and algorithms, and address the following questions: 
\squishlist
\item \textbf{Q1}: Is there a superior core model for multiplex networks? Or depends on the data different core models are needed? (see \autoref{tab:results})
\item \textbf{Q2}: Is there a trade-off between efficiency and cohesiveness? (see \autoref{tab:results}, last part)
\item \textbf{Q3}: How the dimension of degree summery ($d$) affect time and density? (see \autoref{fig:effect-d}) 
\item \textbf{Q4}: How do our algorithms scale with respect to the graph size? (see \autoref{fig:scale})
\item \textbf{Q5}: How does our approximation algorithm perform compare to the optimal solution and baselines? (see \autoref{fig:approx})
\item \textbf{Q6}: How well our mathematical model of user engagement can predict the users' departure from the network? (see \autoref{fig:user-engagement})
\item \textbf{Q7}: Do we need multiple games to model user engagement? (see \autoref{fig:singlelayer})
\squishend
Additional experiments and the details of setups are in Appendix~\ref{app:experimental-setup}.

\head{Datasets}
We perform experimental evaluation on thirteen real networks (ten datasets for evaluating the algorithms and 3 datasets for validating the user engagement model) including social~\cite{Friendfeed, Google+, Twitter_datasets}, genetic~\cite{homo}, co-authorship~\cite{FirmTruss}, financial~\cite{FAO}, and co-purchasing networks~\cite{amazon_datset}, whose main characteristics are summarized in Table~\ref{tab:results}. The detailed description of datasets is in Appendix~\ref{app:datasets}

\head{Baselines}
\score{} is a unifying family of dense structures and so existing dense subgraph models are special cases of \score s. We use state-of-the-art algorithms FirmCore~\cite{FirmCore} and ML $\mathbf{k}$-core~\cite{MLcore}, as well as other seven variants of \score s with different summarizer $\texttt{S}(.)$. WFirmCore is introduced in \S~\ref{sec:WFC}. $\texttt{S}(.) = \textsc{Sum}(.)$ and $= \textsc{WSum}(.)$ use the summation and weighted summation of elements in the degree vector, respectively. $\texttt{S}(.) = \textsc{MGcn}(.)$ and $= \textsc{MGAT}(.)$ use \textsc{Gcn}~\cite{gcn} and \textsc{Gat}~\cite{gat} variants of the multiplex graph neural network in \cite{anomuly}, respectively, to encode degree vectors. Finally, $\texttt{S}(.) = \textsc{Stat}(.)$ uses statistical inference discussed in \S~\ref{sec:proper-s}.

\head{Cohesiveness} 
We first, compere the the density of different variants of \score s (including existing state-of-the-art family of dense subgraphs in the literature) using three different existing density measures for multiplex networks, i.e., edge density (clique density), ML degree density~\cite{MLcore}, and our proposed density in Problem~\ref{Problem:density}. The results are reported in Table~\ref{tab:results}. \circled{1} The results show that there is no single core model that fits all datasets as the network topology and the distribution of node degrees in different layers are the main indicators of what summarizer function is the best representative of nodes neighborhood. The proposed WFirmCore, however, shows promising overall performance compared to other variants, more specifically with respect to edge density and our density. The main reason is its flexibility to consider different layer weights with respect to different nodes. \circled{2} The only exception is in genetic networks, where learning-based summarizer functions, i.e., \textsc{MGcn} and \textsc{MGat} achieve the best overall results in three density measures, possibly due to complex interactions of entities. \circled{3} Learning-based and statistical inference-based methods consistently achieve high density with respect to all density measures. The main reason is that these methods do not use a pre-defined rule/pattern and find a good summarizer in data-driven manner. \circled{4} As expected, (weighted) collapsing the multiplex networks, by using $\texttt{S}(.) = \textsc{Sum}(.)$ and $= \textsc{WSum}(.)$ achieve poor performance as it misses complex interactions in different layers, causing impossibility of inference about nodes' neighborhood (Lemma~\ref{lemma:multiplex-vs-simple}). \circled{5} The superior performance of WFirmCore over FirmCore shows the importance of considering layer weights when finding dense structures.

\setcounter{figure}{4}
\begin{figure*}
\begin{minipage}[c]{0.64\textwidth}
\centering
% \hspace*{-5ex}
        \includegraphics[width=0.44\textwidth]{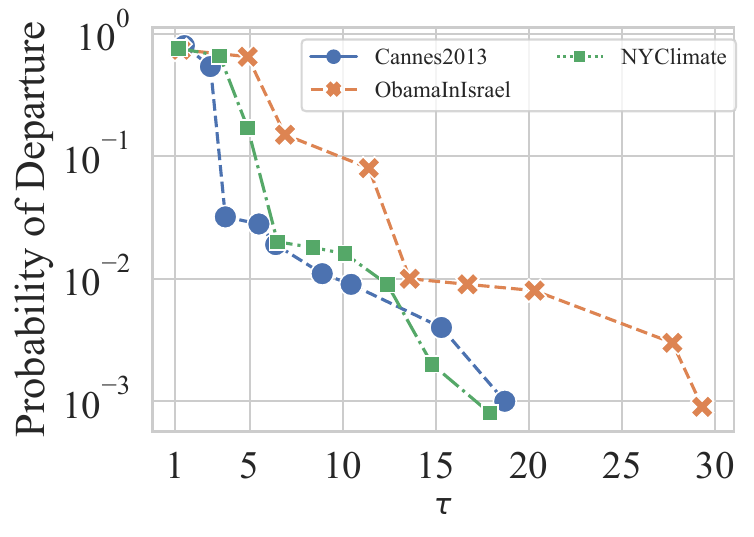}~
    \includegraphics[width=0.47\textwidth]{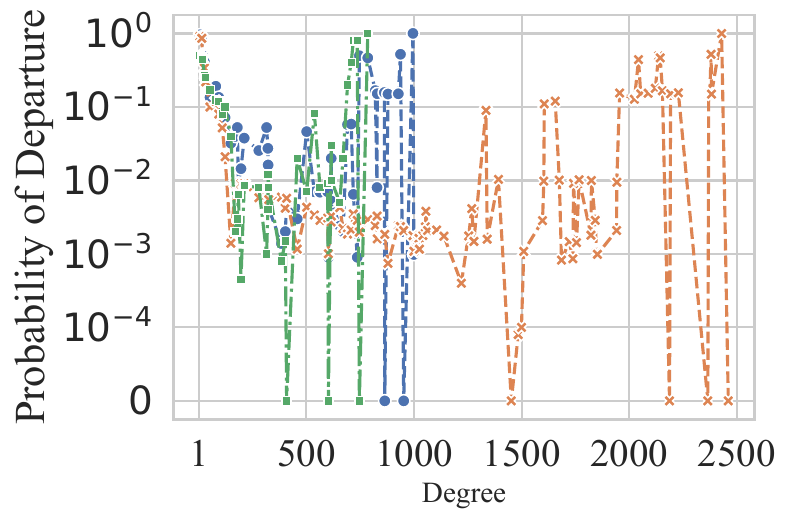}
    \vspace{-3ex}
    \captionof{figure}{Probability of departure vs. nodes' engagement level $\tau$ (Left), and nodes' average degree (Right).}
    \label{fig:user-engagement}
\end{minipage}~\hfill
\begin{minipage}[c]{0.33\textwidth}
\centering
\vspace{-2ex}
\includegraphics[width=0.84\linewidth]{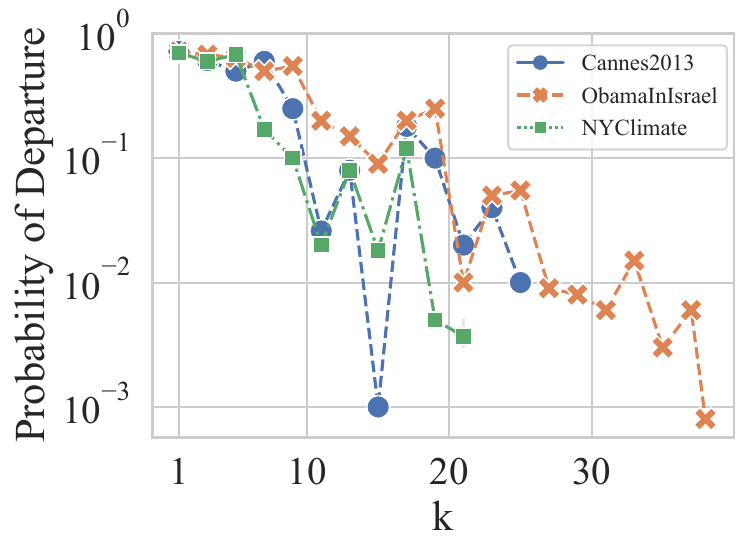}
\vspace{-3ex}
    \captionof{figure}{Probability of departure vs. core number in the collapsed graph.}
    \label{fig:singlelayer}
    \vspace{-2ex}
\end{minipage}
\end{figure*}

\setcounter{figure}{3}
\begin{figure}
    \hspace{-0.4cm}
    \includegraphics[width=0.45\textwidth]{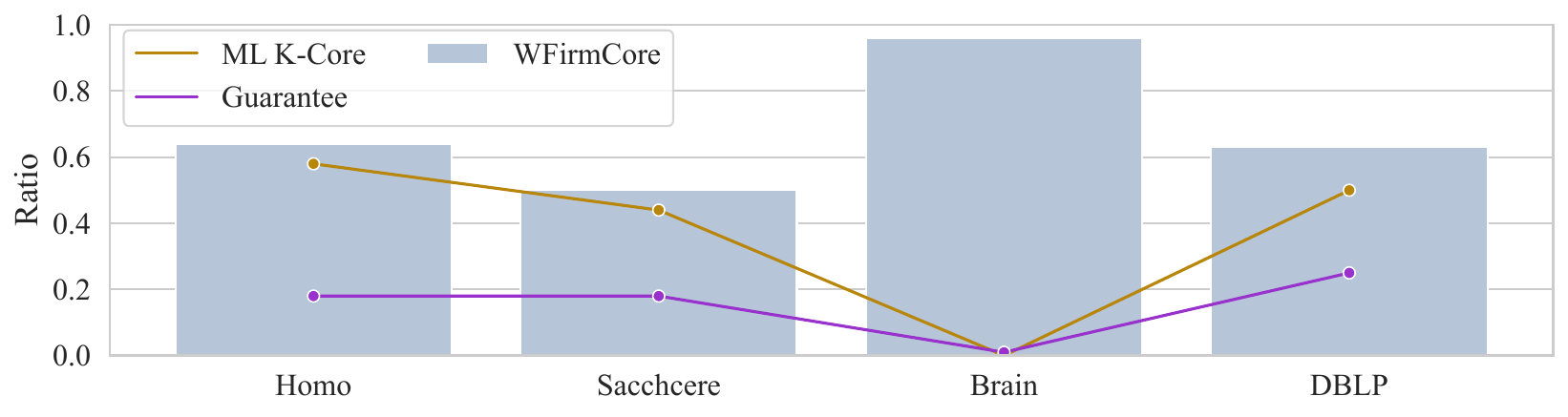}    
    \vspace{-2ex}
    \caption{The empirical approximation factor of WFC-Approx.}
    \label{fig:approx}
    \vspace{-3ex}
\end{figure}

\setcounter{figure}{6}

\head{Efficiency of Algorithms}
The last part of Table~\ref{tab:results} compares the running time of the different dense subgraph models. As expected, using simple summarizer function (e.g., $\textsc{Sum}(.)$ and Top-$\lambda(.)$) to summarize the degree vector to a single number results in more efficient decomposition algorithms. Furthermore, it shows that using high dimensional representation of degree (i.e., large $d$) makes the approach impractical for networks with either large size or $|L|$.

\head{Conclusion of Table~\ref{tab:results}}
Table~\ref{tab:results} is a clear evidence of the three-way trade-off of efficiency (both time and memory), effectiveness (i.e., density), and too hard degree constraints (the output dimension of $\texttt{S}(.)$) in real-world multiplex networks. That is, for large networks (either graph size or \#layers) we have to use methods that summarize the degree vector to a single number. The main reason of this three-way trade-off is that there is no pre-defined pattern/rule that fits all networks, and so we need to decide about the patterns of interests in a data driven manner, depending on the network topology and available computation capacity.

\head{The effect of $d$}
To evaluate the effect of $d$ on both running time and density of \score s, we vary its value and report the density and running time of \score on a subset of StackOverflow in Figure~\ref{fig:effect-d}. As expected, larger $d$ always results in a slower decomposition algorithm, as we have larger search space. On the other hand, however, finding the best value of $d$ to achieve higher density depends on the network topology and the trend can vary from a dataset to another. Accordingly, increasing $d$ might result in lower density as it might be a too hard constraint for core structures.

\head{Scalability}
Figure~\ref{fig:scale} demonstrates the effect of \#layers and graph size on the running time of the \score{} decomposition algorithm. In this part, we use different versions of a variable size subgraph of StackOverflow obtained by selecting a variable number of layers from 1 to 24. The running time of \score, scales gracefully with respect to both $|E|$ and $|L|$ and also scales near linear with respect to $|L|$. Notably, based on the running time results in Table~\ref{tab:results}, some variants of \score{} (e.g., FirmCore~\cite{FirmCore} and WFirmCore) can scale to graphs with billions of edges.

% \head{Data-driven Dense Structures}

\head{Densest Subgraph: Approximation Performance}
To empirically evaluate the quality of the solution found by our proposed approximation algorithm for Problem~\ref{Problem:density}, we report the ratio of the density of the found solution over the optimal density in Figure~\ref{fig:approx}. The results show that WFC-Approx algorithm, in practice, finds solutions with higher density than its guarantee. We further compare it with ML $\mathbf{k}$-core, which is the state-of-the-art algorithm for approximating the densest subgraph problem. The results show the superior performance of our algorithm.

\head{DBLP Case Study}
Existing density measure forces all nodes to exhibit their high degree in a fixed subset of layers. To support our motivation of designing a new density measure that allows nodes to exhibit their high degree in flexible subsets of layers,  we perform a case study on DBLP dataset. Here, each node is a researcher and two nodes are connected if they have published a paper with each other. The type of connections are obtained using LDA~\cite{LDA} algorithm on the topics and abstracts of the paper. Accordingly, each layer is the collaboration network in a specific research topic. We then find and compare the densest subgraph by our approach and \citet{MLcore}. The found subgraph by our approach consists of two communities, $\mathcal{A}$ and $\mathcal{B}$, with edge density of 0.41, while the found densest subgraph by \citet{MLcore} is only $\mathcal{A}$ with edge density of 0.38. Since the ML density forces all nodes to have high degree in a fixed set of layers, it misses the $\mathcal{B}$ community as they collaborated in different topics than the $\mathcal{A}$ community. The found communities are visualized in Appendix~\ref{app:DBLP}.

\head{Modeling User Engagement}
In this experiment, we evaluate the correlation between the proposed user engagement level, $\tau(.)$, and the probability of departure from the network. We use three temporal multiplex networks obtained from $\mathbf{X}$ (formerly Twitter) during exceptional events with 3 layers, corresponding to repost, mentions, and replies between users (details are in Appendix~\ref{app:datasets}). We consider a user departed if they stop posting about the topic. Figure~\ref{fig:user-engagement} (Left) reports the probability of departure with respect to $\tau(.)$. Interestingly, most users with a low (resp. high) engagement level depart (resp. stay in) the network. These results support our mathematical formulation for user engagement. Figure~\ref{fig:user-engagement} (Right) reports the same experiments with respect to the average degree. The results show that a simple average degree is not a good indicator of engagement in social networks.

\head{Importance of Multiple Games}
Figure ~\ref{fig:singlelayer} reports the probability of departure with respect to the core number in a collapsed graph obtained from a multilayer graph by merging edges. The results indicate that the core number in a collapsed network, while information about different types of connections is available, is not a good indicator of engagement in social networks.

\section{Conclusion}
We present a new family of dense subgraphs in multiplex networks that unifies existing families using a single function $\texttt{S}(.)$. We show that \score{} has the nice properties of $k$-cores in simple graphs and suggest three methods (i.e., statistical inference-, sampling-, and learning-based) to choose function $\texttt{S}(.)$ in a data-driven manner. We further propose a new density measure for multiplex networks, and design a new variant of \score s to effectively approximate the solution of the densest subgraph problem. Finally, based on \score s, we propose a new mathematical model for modeling user engagement in social networks with different types of interactions. Our experimental evaluation shows the efficiency and effectiveness of
our algorithms and supports the proposed mathematical model of user engagement.

\newpage
\bibliographystyle{ACM-Reference-Format}
\bibliography{main}

\appendix

\section{Notations}\label{app:notation}
We provide Table~\ref{tab:notations} to summarize all the used notations through the paper:

\begin{table}[h]
    \centering
     \caption{Notations through the paper.}
    \vspace{-1ex}
    \resizebox{1\linewidth}{!}{
    \begin{tabular}{c|c}
    \toprule
        Notation &  Meaning \\
         \midrule
         \midrule
        $V$ & The set of all vertices. \\
        $L$ &  The set of all layers.\\
        $E \subseteq V \times V \times L$ & The set of all connections. \\
        \multirow{1}{*}{$\w$} &  The weight function that assigns\\
        & a weight to each pair of nodes and layers.\\
        \multirow{1}{*}{$G = (V, E, L, \w)$} & A multiplex network with node, edge, \\
        & and layer sets $V, E$, and $L$.\\
        $N_\ell(u)$& The set of $u$'s neighbors in layer $\ell$. \\
        $\deg(u)$ & The degree vector of $u \in V$. \\
        $\deg_{\ell}(u)$ &  The degree of $u \in V$ in layer $\ell \in L$. \\
        $\deg_{\ell}^{H}(u)$ &  The degree of $u \in V$ within subgraph $H$ in layer $\ell \in L$. \\
        $\rho(S)$ & Density of Subgraph $S$ (Problem~\ref{Problem:density}). \\
        $\Omega$& The maximum summation of layer weights over a node. \\
        $\w^*$ & The summation of all layer weights.\\
        $S_{\text{SL}}$  &Densest single-layer subgraph among all layers.  \\
        $\mu^*$ & Minimum degree of the  densest single-layer subgraph ($S_{\text{SL}}$).\\ 
        $\psi$ & Maximum value that $(\mu^*, \psi)$-WFirmCore is non-empty \\
        $\tau(u)$ & The u's engagement level (Def.~\ref{dfn:engagement}).\\
    \bottomrule
    \end{tabular}
    }
    \label{tab:notations}
\end{table}

\section{Backgrounds}\label{app:backgrounds}

\head{ML $\mathbf{k}$-core Decomposition}
Since in multiplex networks, each node degree is a vector, \citet{azimi-etal} suggest to use different coreness number for each layer and define the multiplex $\mathbf{k}$-core as follows: 

\begin{dfn}[Multilayer $\mathbf{k}$-core]\cite{azimi-etal}
    Given an unweighted undirected multiplex graph $G = (V, E, L)$, and an integer vector $\mathbf{k} = [k_{i}]_{1 \leq i\leq |L|}$, Multilayer $\mathbf{k}$-core is a maximal subgraph $C_{\mathbf{k}}$ such that for each node $v \in C_{\mathbf{k}}: \:\: \deg^{C_{\mathbf{k}}}_{\ell_i}(v) \geq k_i$.
\end{dfn}
\noindent
\citet{MLcore} presents the first core decomposition algorithms of multiplex networks using the above definition of cores. Since the number of possible vectors $\mathbf{k}$ is $\mathcal{O}(d_{\max}^{|L|})$, where $d_{\max}$ is the maximum degree in the network and $|L|$ is the number of layers, any algorithm that finds all possible multiplex $\mathbf{k}$-cores of a multiplex graph should have exponential time complexity. To this end, the authors focus on optimizing and improving the implementation of exponential-time algorithm that performs brute force search to find all possible $\mathbf{k}$-cores.

\begin{figure}
    \centering
    \includegraphics[width=0.65\linewidth]{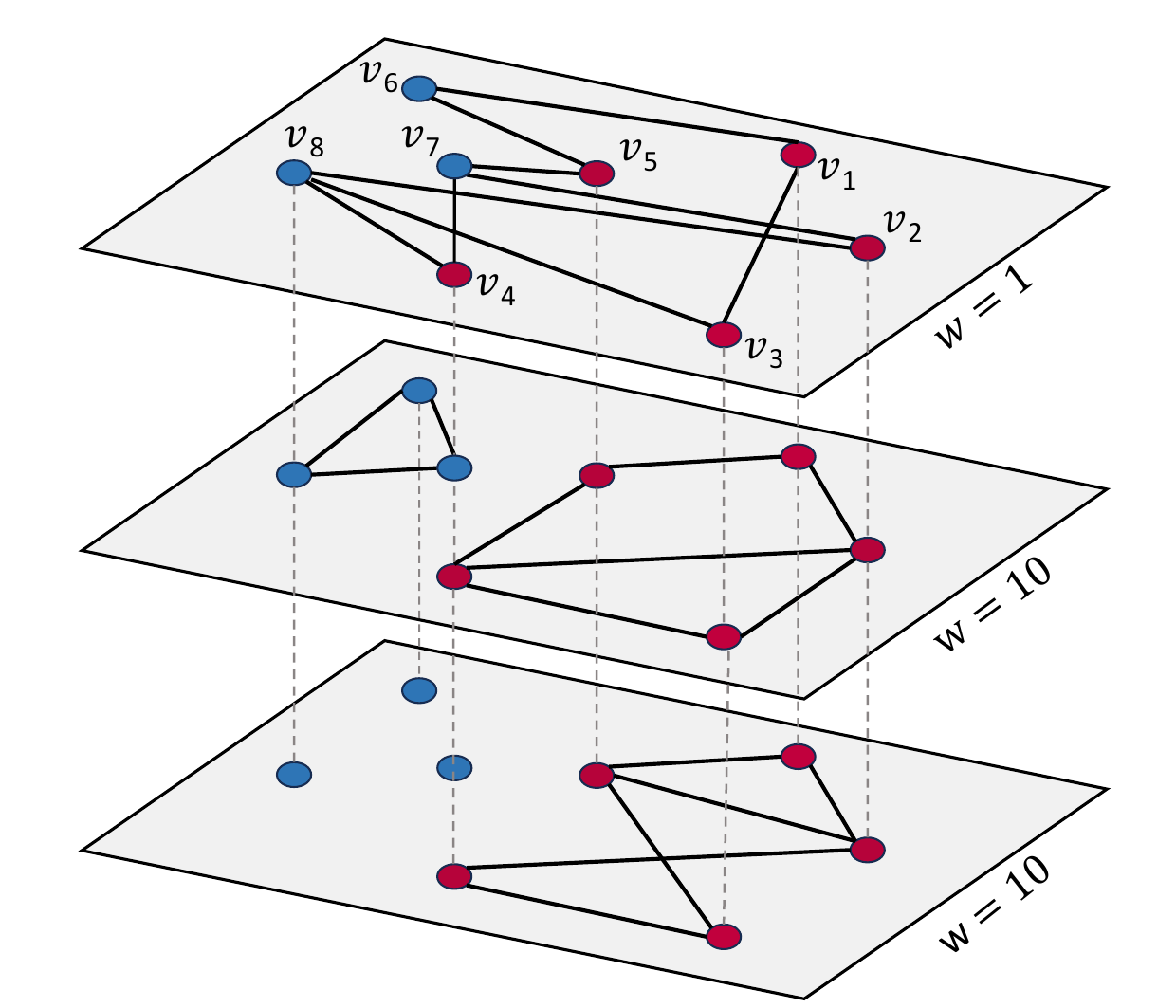}
    \caption{An example of multiplex collaboration network.}
    \label{fig:example2}
    \vspace{-2ex}
\end{figure}

\noindent
The main drawback of their core decomposition algorithm is its \emph{exponential running time complexity in the number of layers}, which makes it prohibitive for large graphs even with a small number of layers. Another drawback of their approach is that they assume that the importance of each layer is the same. However, in multiplex networks, some layers for each node might be noisy, insignificant, or incomplete.

\head{FirmCore Decomposition}
To address the time complexity of the ML $\mathbf{k}$-core, \citet{FirmCore} present a new concept of core in multiplex networks that each node has Top-$\lambda$ degree at least $k$ within the subgraph. Formally they define FirmCore as follows:

\begin{dfn}[FirmCore]\cite{FirmCore}
Given an unweighted undirected multiplex graph $G = (V, E, L)$, an integer threshold $1 \leq  \lambda \leq |L|$, and an integer $k \geq 0$, the $(k, \lambda)$-FirmCore of $G$ ($(k, \lambda)$-FC for short) is a maximal subgraph $H = G[C_{k}] = (C_{k}, E[C_{k}], L)$ such that for each node $v\in C_k$ there are at least $\lambda$ layers $\{\ell_1, ..., \ell_\lambda\} \subseteq L$ such that $\text{deg}^H_{\ell_i}(v) \geq k$, $1\leq i\leq \lambda$.
\end{dfn}

\noindent
The main drawback of this definition is that, it assumes that the importance of each layer for each node is the same.
d

\head{Densest Common Subgraph Problem}
Jethava and Beerenwinkel \cite{densest-common-subgraph} introduced the concept of the densest common subgraph problem, which is to find a subgraph that maximizes the minimum average degree across all input layers in the graph. They devised a linear-programming formulation and a greedy heuristic to address it. More formally they define density as follows:

\begin{dfn}[Common Subgraph Density]\label{dfn:common-density}
    Given a mutliplex network $G = (V, E, L)$, the common subgraph density of $G[H]$ is defined as:
    \begin{equation}
        \rho(H) = \min_{\ell \in L} \frac{|E_\ell[H]|}{|H|}.
    \end{equation}
\end{dfn}

\section{Additional Related Work}\label{app:rw}

\head{Problem of Densest Subgraph}
For additional discussion on the densest subgraph problem, we refer to the recent survey by \citet{densest-survey}.

\head{Heterogeneous Networks}
ML networks can be seen as a type of heterogeneous information networks (HINs). But the two definitions are being used for slightly different meanings. ML graphs emphasize multiple types of relationships between similar types of entities. On the other hand, HINs emphasize heterogeneous types of entities connected
by different relationships. Fang et al.~\cite{HIN-core} define the core structure of heterogeneous information networks. Liu et al.~\cite{HINliu} define the core structure of bipartite graphs, and Zhou et al.~\cite{HIN2}extend it to $k$-partite networks. However, all these models emphasize different types of entities connected by different types of relations, which differs from the concept of ML networks. Consequently, their approach is not applicable to ML networks.

\head{Cores in Temporal Networks}
Wu et al. ~\cite{core-temporal} extend the notion of core to temporal graphs. While their definition does not directly consider any temporal peculiarity and can apply to ML networks, it is equivalent to collapsing the layers, removing edges that occur less than a threshold, and finding cores in the resulting single-layer graph. While it focuses on the frequency of interactions and conceptually is appropriate for temporal graphs, it cannot capture complex relationships in ML networks. However, temporal graphs can be seen as special cases of multiplex networks, where each layer is a snapshot of the network.

\section{Detailed Examples}

\noindent
\textcolor{c1}{\textbf{How Layer Weight Can Affect the Core Structures?}}
Figure \ref{fig:example2} shows a multiplex network, where $\w(v,l_1) = 1$, $\w(v,l_2) = 10$, and $\w(v,l_3) = 10$,  $\forall v \in V$. FirmCore decomposition treats all relation types equally and ignores the weights of the layers, therefore causing missing information. For example, in the multilayer graph depicted in Figure \ref{fig:example2}, FirmCore decomposition, while ignoring weights, finds $\{v_1, v_2, ..., v_8\}$ (the union of blue and red nodes) as a $(1,1), (2,1), (1,2), (2,2)$-FirmCore and is not able to find any other subgraph. However, based on weights, the second and third layers are more important. Therefore, the WFirmCore decomposition finds $\{v_1, v_2, ..., v_5\}$ (red nodes) as a $(2,20)$-WFirmCore and $\{v_1, v_2, ..., v_8\}$ (the union of blue and red nodes) as a $(1,11), (2,11)$-WFirmCore. This means that WFirmCore takes into account the importance of layers for each node and provides a more diverse solution space.

%\head{Toy Example for Algorithm~\ref{alg:dfs-path}}

%\head{Toy Example for Algorithm~\ref{alg:S-decomposition}}

%\head{Toy Example for Algorithm~\ref{alg:WFirmCore}}

%\head{Toy Example for Algorithm~\ref{alg:Approx1}}

\section{Proofs}\label{app:proofs}

\subsection{Lemma~\ref{lemma:special-case}}

\begin{proof}
First, let $\w = \mathbf{1}$ and $S(\mathbf{X})= \mathbf{X}$, then $\mathbf{k}$-\score{} is equivalent to multilayer $\mathbf{k}$-core~\cite{MLcore}. 

Also, 
 given a subset of layers $\hat{L} \subseteq L$, let $\w = \mathbf{1}$ and $S(\mathbf{X}) = \mathbf{X}_{\hat{L}}$ (corresponding elements to $\hat{L}$), and $\mathbf{k} = [k, k, \dots, k]_{1 \times |\hat{L}|}$ then $\mathbf{k}$-\score{} is equivalent to CoreCube~\cite{CoreCube}. 

Moreover, given $\lambda \in \mathbb{N}$, let $S(\mathbf{X}) = \text{Top-}\lambda(\mathbf{X})$, then $\mathbf{k}$-\score{} is equivalent to FirmCore~\cite{FirmCore}.     
\end{proof}

\subsection{Proposition~\ref{prop:maximal-scv}}
\begin{proof}
    Since all subgraphs are $\mathbf{k} = [0]_{1 \times d}$-\score, the set of all coreness vector of each subgraph is non-empty and so the maximal coreness vector exists. To show that this maximal coreness vector is unique, we use contradiction. Assume that there are two $\mathbf{k} = [k_{i}]_{1 \leq i\leq d} \neq \mathbf{k}' = [k'_{i}]_{1 \leq i\leq d}$ such that both are maximal coreness vector of $G[C]$. Based on the definition of maximal coreness vector, there are indices $\ell_1$ and $\ell_2$ such that $k_{\ell_1} > k'_{\ell_1}$ and $k'_{\ell_2} > k_{\ell_2}$. Now define $\tilde{\mathbf{k}} = [\tilde{k}_{i}]_{1 \leq i\leq d}$ such that $\tilde{k}_{\ell} = \max\{k_\ell, k'_{\ell}\}$ for all $1\leq \ell \leq d$. Based on the definition of maximal coreness vector, $\tilde{\mathbf{k}}$ is also coreness vector of $G[C]$, which contradicts the maximality of $\mathbf{k} = [k_{i}]_{1 \leq i\leq d} \neq \mathbf{k}' = [k'_{i}]_{1 \leq i\leq d}$. Therefore, the maximal coreness vector for each \score{} exists and is unique.
\end{proof}

\subsection{Proposition~\ref{prop:unique-core}}
\begin{proof}
    Suppose that $G[C_{\mathbf{k}}]$ and $G[C'_{\mathbf{k}}]$ are two distinct $\mathbf{k}$-\score s of $G$. We know that $G[C_{\mathbf{k}}]$ is a maximal subgraph such that $ \forall v \in G[C_{\mathbf{k}}]: \:\: \texttt{S}(\deg^{C_{\mathbf{k}}}(v))_{i} \geq k_i$ for all $1 \leq i \leq d$. Similarly, $G[C'_{\mathbf{k}}]$ is a maximal subgraph with the same property. Then any node $v \in G[C_{\mathbf{k}} \cup C'_{\mathbf{k}}]$ has been in either $G[C_{\mathbf{k}}]$ or $G[C'_{\mathbf{k}}]$, which means that $\texttt{S}(\deg^{C_{\mathbf{k}} \cup C'_{\mathbf{k}}}(v))_{i} \geq \texttt{S}(\deg^{C_{\mathbf{k}}}(v))_{i} \geq k_i$. Therefore, $G[C_{\mathbf{k}} \cup C'_{\mathbf{k}}]$ satisfies the \score{} conditions, contradicting the maximality of $G[C_{\mathbf{k}}]$ and $G[C'_{\mathbf{k}}]$.
\end{proof}

\subsection{Proposition~\ref{prop:hierarchical1}}
\begin{proof}
     Let $v \in G[C_{\mathbf{k}}]$, based on the definition of \score s, we have: $\texttt{S}(\deg^{C_{\mathbf{k}}}(v)) \geq \mathbf{k}$ and so $\texttt{S}(\deg^{C_{\mathbf{k}}}(v)) \geq \mathbf{k}'$. This implies that $v \in G[C_{\mathbf{k}'}]$ and so $G[C_{\mathbf{k}}] \subseteq G[C_{\mathbf{k}'}]$.
 \end{proof}

\subsection{Theorem~\ref{thm:unique-core}}
\begin{proof}
    First, it is clear that $G[C]$ is $([\min_{u \in C}\texttt{S}(\deg^{C}(u))_i]_{i = 1}^d)$-\score, since for each node $u \in V$ and $i = 1, \dots, d$ we have:
    \begin{align}
        \texttt{S}(\deg^{C}(u))_i \geq \min_{u \in C}\texttt{S}(\deg^{C}(u))_i,
    \end{align}
    which satisfies the condition of the \score s. Now let $\hat{\mathbf{k}} = [\hat{k}_{i}]_{i = 1}^{d}$ be the maximal \texttt{SCV} index of $G[C]$. This means that there is index $j$ such that $\hat{k}_j > \min_{u \in C}\texttt{S}(\deg^{C}(u))_j$. This contradicts the definition of \score s as the node that attains the $\min_{u \in C}\texttt{S}(\deg^{C}(u))_j$ cannot satisfy \score conditions for vector $\hat{\mathbf{k}} = [\hat{k}_{i}]_{i = 1}^{d}$.
\end{proof}

 \subsection{Lemma~\ref{lemma:multiplex-vs-simple}}
 \begin{proof}
     We use random graphs based on Erdős–Rényi model. Let $G = (V, E, L)$ be a multiplex graph, where each $G_\ell$ is randomly generated by Erdős–Rényi model and is independent of other $G_{\ell'}$s, and $G_{\text{SL}}$ be its single-layer simple graph representation (without multiple connections between pairs of nodes) by aggregating all connections in different relation types. Given an arbitrary vertex $u \in V$, we know that the degree distribution of $u$ in $G_\ell$ follows binomial distribution. Accordingly, here, we see each degree vector $\deg(u)$ as $|L|$ i.i.d. samples $\deg_1(u), \dots, \deg_{|L|}(u)$ from binomial distribution. Our goal is to estimate parameter $\theta_u$ as the summary of $u$'s degree vector for each $u \in V$ such that $\theta_u$ carries all the information about all $\deg_\ell(u)$. To this end, we need to have sufficient statistics for binomial distribution, which is $\hat{\theta}_u = \sum_{i = 1}^{|L|} \deg_{i}(u)$. Therefore, $\hat{\theta}_u$ carries all the information we need to inference about degree vector of $u$. On the other hand, however, the degree of $u$ in the aggregated graph is:
\begin{align}
    \deg_{\text{SL}}(u) = \sum_{i = 1}^{|L|} \deg_{i}(u) - \:\: R(u),
\end{align}
where $R(u)$ is the number of $u$'s redundant connections, i.e.,  
\begin{align*}
    R(u) = |\{(u, v)| \exists \ell_1, \ell_2 \in L \:\: \text{such that} \:\:\: (u, v, \ell_1), (u, v, \ell_2) \in E \}|.
\end{align*} 
It is simple to see that $\mathbb{E}[R(u)]\neq 0$ (note that $R(u) \geq 0$) and so
\begin{align}
    \mathbb{E}[\deg_{\text{SL}}(u)] = \mathbb{E}[\sum_{i = 1}^{|L|} \deg_{i}(u)] - \mathbb{E}[R(u)] \neq \mathbb{E}[\sum_{i = 1}^{|L|} \deg_{i}(u)].
\end{align}
Since $\sum_{i = 1}^{|L|} \deg_{i}(u)$ is minimal sufficient, it is simple to see that $\sum_{i = 1}^{|L|} \deg_{i}(u) - \:\: R(u)$ is not a sufficient statistics and so does not carry all the information we need to inference about degree vector of $u$. 
 \end{proof}

\subsection{Lemma~\ref{lemma:span-core}}

\begin{proof}
    Given $\lambda$, let $\w_\ell = 2^\ell$ for all $\ell \in L$ and $\texttt{S}(\mathbf{X}) = \text{Top-}\lambda(\mathbf{X})$. Note that if we consider the binary representation of $\lambda$, then by finding $(k, \lambda)$-WFirmCore we exactly know what layers are choosen. Accordingly, given $\Delta$ in span core, one can consider the set of all values of $\Delta$ such that the summation of layer weights for all possible $\Delta$ consecutive layer is calculated. Let us call this set $\Phi$. Now funding $(k, \lambda)$-WFirmCore for $\lambda \in \Phi$ is equivalent to span core with parameter $\Delta$.
\end{proof}

\subsection{Theorem~\ref{thm:WFirmCore-hardness}}

\begin{proof}
    To show the NP-hardness of WFirmCore decomposition, we show that its special case is NP-hard. To this end, we assume that the weight of each layer is the same across different nodes, i.e., $\w(u, \ell) = \w(v, \ell)$ for any $u, v \in V$ and $\ell \in L$. Given a sequence of layer weights $w_1, w_2, \dots, w_{|L|}$,  the decision problem of whether there is a non-empty $(k, \lambda)$-WFirmCore can be simply reduced to the well-known NP-hard problem of the \emph{Subset Sum} over $w_1, w_2, \dots, w_{|L|}$, as its YES (resp. NO) instance means there is (resp. is not) a subset of $w_i$s with summation of~$\lambda$.

    Since the weight of each layer is integer, we copy each layer by its weight. Therefore, the resulted multiplex graph is unweighted. Now, layer-weighted FirmCore on the original graph is equivalent to simple FirmCore on the constructed graph and since FirmCore decomposition has polynomial time complexity, i.e., $\mathcal{O}(|E||\tilde{L}|^2 + |V||\tilde{L}|\log |\tilde{L}|)$), WFirmCore decomposition can be done in Pseudo-polynomial time. Note that here $\tilde{L}$ is the summation of all layer weights.
\end{proof}

\subsection{Theorem~\ref{thm:MDS-hardness}}

\begin{proof}
    Assume the simple case where $\w(u, \ell) = 1$ for all $u \in V$ and $\ell \in L$. We reduce our problem to the problem of densest common subgraph~\cite{densest-common-subgraph}, which is proved to be NP-hard and also cannot have approximation algorithm with a constant factor~\cite{Hardness}. Without loss of generality we can assume that there is no node with degree zero in at least one layer. This is a valid assumption as we can simple add a dummy node and connected to all existing nodes. Therefore, the dummy node is always in the densest subgraph, and removing it cannot affect the subgraph that attains the maximum density.
    Let $G = (V, E, L)$ be an arbitrary instance of the problem of densest common subgraph, let $\beta$  be a large enough number that forces the objective function to choose all the layers. In this case, the objective function is equivalent to the densest common subgraph problem in \cite{densest-common-subgraph}, which is proved to be NP-hard~\cite{Hardness}. 
\end{proof}

\subsection{Theorem~\ref{thm:approximation-factor}}
\begin{proof}
    For any $u \in S$, we let $\Delta_u(S) = |S|\rho(S) - (|S| - 1) \rho(S\setminus \{u\})$. It shows how removing node $u$ can affect the numerator of density function (note that removing any node affect the denominator in the same way). Now, let $S^*$ be the optimal solution with density $\frac{\gamma}{|S^*|}$, removing a node cannot increase the density so we have:
    \begin{align}
        \frac{\gamma}{|S^*|} \geq \frac{\gamma - \Delta_u(S^*)}{|S^*| - 1} \Rightarrow \Delta_u(S^*) \geq \frac{\gamma}{|S^*|} = \rho(S^*).
    \end{align}
    Now note that:
    \begin{align*}
        \Delta_u(S) \leq \max_{\hat{L} \subseteq L} \min_{\ell \in \hat{L}} \deg_\ell(u) |L|^\beta + \sum_{v \in N(v)} |L|^\beta \leq 2|L|^\beta \min_{\ell \in L} \deg_\ell(u).
    \end{align*}
    Now, let $u = \arg\min_{v \in S^*} \min \min_{\ell \in L} \deg_\ell(v)$, it is simple to see that $(\min_{\ell \in L} \deg_\ell(u), 1)$-WFirmCore is not empty and also is considered in Algorithm~\ref{alg:Approx1}, which proves the theorem.
\end{proof}

\subsection{Proposition~\ref{prop:density-comparison}}
\begin{proof}
    This inequality comes from the definition of the objective functions. Let $L^*$ be the selected set of layers in objective of \autoref{eq:pre-density}, then we have:
    \begin{align}
        \rho(G[H]) &= \frac{1}{|S|} \sum_{u \in H} \max_{\hat{L} \subseteq L} \min_{\ell \in \hat{L}} \deg_\ell(u) \times |\hat{L}|^\beta \\
        &\geq \frac{1}{|S|} \sum_{u \in H} \min_{\ell \in L^*} \deg_\ell(u) \times |L^*|^\beta \\
        &= \rho^*(G[H]).
    \end{align}
\end{proof}

\subsection{Theorem~\ref{thm:equilibrium}}

\begin{proof}
    We first show that $\mathbf{k}$-\score{} is an equilibrium. To this end, assume that a user $u$ who has decided to drop out wants to change their decision. In this case, its new utility in each layer will be negative. On the other hand, if a user $u$ who has decided to remain engage wants to change their decision and drop out, then its utility in each layer will be zero. Accordingly, no user wants to change their decision. Now, note that since the definition of $\mathbf{k}$-\score{} is the maximal subgraph with this property, the maximal equilibrium should be the maximal $\mathbf{k}$-\score{}, which we show is unique. 
\end{proof}

\section{How To Choose $\texttt{S}(.)$?}\label{app:prop-S}

\noindent
Given a multiplex network $G$, one might ask ``why does multiplex core structures of $G$ can provide richer information than simple $k$-core~\cite{k-core-first} in its collapsed simple graph representation?''. Using the above approach, we answer this question as follows:

\begin{lemma}\label{lemma:multiplex-vs-simple}
    There are infinitely many multiplex graphs like $G$ such that collapsing it into a simple network (i.e., ignoring the type of connections) causes impossibility of inference about degree vectors.
\end{lemma}

\begin{example}
    It is known that many real-world networks are scale-free and so their degree distribution follows a power law distribution. Given a multiplex network with independent layers, where each layer is scale-free, then since $\texttt{S}(u) = \sum_{i = 1}^{|L|} \deg_{i}(u)$ is a sufficient statistics for power law distribution, the output of the above procedure is function $\texttt{S}(.)$.
\end{example}

\begin{example}
        Given a multiplex network $G$, where the degree distribution in each layer is uniform, $(k, 1)$-FirmCore has all the information we need to inference about vertices degree. The reason is, $(k, 1)$-FirmCore is the maximal subgraph, where each node has Top-1 degree of at least $k$ and Top-1 is sufficient statistics for uniform distribution. 
\end{example}

\section{Experimental Setup}\label{app:experimental-setup}
All algorithms are implemented in Python and compiled by Cython. The experiments are performed on a Linux machine with Intel Xeon 2.6 GHz CPU and 128 GB RAM.
\subsection{Datasets}\label{app:datasets}
We perform extensive experiments on ten real networks~\cite{homo, Twitter_datasets, FAO, dblp, FirmCore, FirmTruss, Friendfeed, amazon_datset, Google+} including social, genetic, co-authorship, financial, brain, and co-purchasing networks, whose main characteristics are summarized in Table 1. 
SacchCere and Homo~\cite{homo} are biological networks concerning different genetic interactions between genes in Homo Sapiens and Saccharomyces Cerevisiae, respectively. FAO~\citep{FAO} represents various trade relationships among countries, where layers signify products, nodes represent countries, and edges at each layer depict import/export connections among countries. Brain~\cite{FirmTruss, CS-MLGCN}  is derived from the functional magnetic resonance imaging (fMRI) of 520 individuals using the same methodology as in \cite{FirmTruss}. In this dataset, each layer represents the brain network of an individual, where nodes correspond to brain regions, and edges show the statistical association between the functionality of these nodes. DBLP~\cite{dblp} is a co-authorship network derived following the methodology in ~\cite{MLcore, FirmTruss}, where each layer represents topics of papers determined using LDA topic modeling~\cite{LDA}. Nodes correspond to researchers, and edges denote co-authorship relationships. Amazon~\cite{amazon_datset} is a co-purchasing network, in which each layer is associated with one of its four snapshots between March and June 2003. FFTwitter \cite{Friendfeed} is a multi-platform social network where layers correspond to interactions on Friendfeed and Twitter, and nodes represent users registered on both platforms. Friendfeed~\cite{Friendfeed} contains interactions such as commenting, liking, and following among its users over two months. StackOverflow represents user interactions from the StackExchange website, where each layer corresponds to interactions during a specific hour of the day. Google+~\cite{Google+, FirmCore} is a billion-scale network comprising four snapshots from Google+ captured between July and October 2011. ObamaInIsrael~\cite{Twitter_datasets} represents retweeting, mentioning, and replying among Twitter users, with a focus on Barack Obama's visit to Israel in 2013. Cannes~\cite{Twitter_datasets} represents retweeting, mentioning, and replying among Twitter users, with a focus on the Cannes Film Festival in 2013. Finally, NYClimate~\cite{Twitter_datasets} represents retweeting, mentioning, and replying among Twitter users, with a focus on the  People's Climate March in 2014. We use an unsupervised approach to automatically assign weight to the datasets that do not have layer weights~\citep{anomuly}.

\subsection{Baselines}\label{baselines}
We compare our defined core structure with state-of-the-art core decompositions in ML networks. ML $\mathbf{k}$-core~\cite{MLcore} uses different coreness numbers for each layer, and for a given $\mathbf{k}$, the ML $\mathbf{k}$-core is defined as a maximal subgraph such that each node in layer $i$ has a degree of at least $k_i$ within the subgraph. On the other hand, FirmCore~\cite{FirmCore} presents a new concept of core in multiplex networks, where each node has a top-$\lambda$ degree of at least $\lambda$ within the subgraph. For finding the densest subgraph based on density definition in \cite{MLcore}, both studies perform the core decomposition and return the found densest subgraph.

\section{Additional Experimental Results}

%\subsection{Running Time Comparison}

% \subsection{Learning $\texttt{S}(.)$}

\subsection{Densest Subgraph: Approximation Performance} %effect of alpha

%\head{The Number of Selected Layers}

\begin{figure}
    \includegraphics[width=0.245\textwidth]{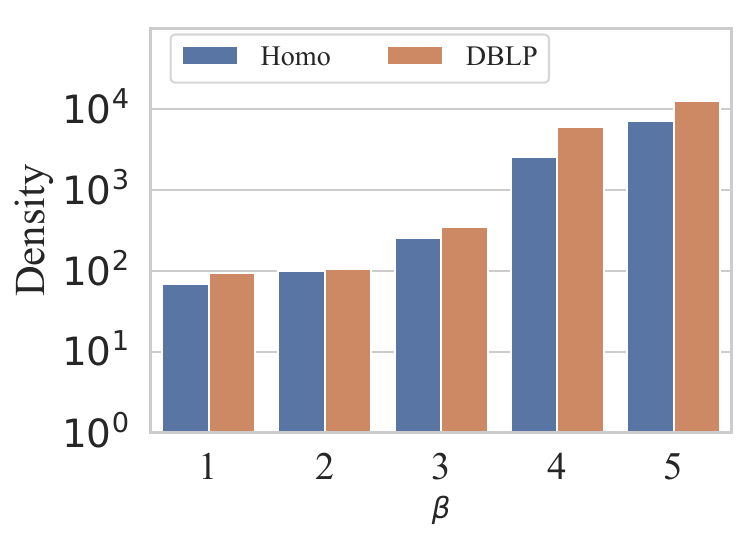}~
    \includegraphics[width=0.23\textwidth]{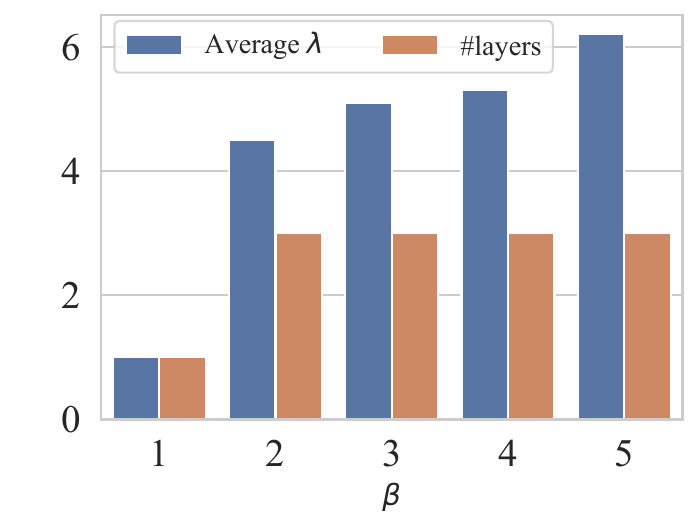}
    \vspace{-3ex}
    \caption{Our density values for the output densest subgraphs with varying $\beta$ (Left),  the number of selected layers in densest subgraphs based on ML density and the average $\lambda$ of densest subgraphs based on our density with varying $\beta$ on Homo dataset (Right).}
    \label{fig:effect-beta}
\end{figure}

\head{The Effect of $\beta$} We experimentally evaluate our densest algorithm using the datasets listed in Table \ref{tab:results}. In Figure \ref{fig:effect-beta} (left), we present the results of our density values for the Homo and DBLP datasets, with varying $\beta$. As expected, the density value increases exponentially as $\beta$ increases.
In addition, in Figure \ref{fig:effect-beta} (right), we evaluate the effect of increasing $\beta$ on the number of selected layers in the output densest subgraph for the Homo dataset based on ML density, as well as the average $\lambda$ in our density. In the ML density function, a fixed number of layers is selected where all nodes exhibit their highest density in these fixed layers. Therefore, the number of selected layers is the same for all nodes. On the other hand, our density function does not impose hard constraints on a fixed number of layers, allowing each node to consider different $\lambda$ values that maximize density. Therefore, we calculate the average $\lambda_i$ over nodes $v_i$. As expected, as $\beta$ increases, both the number of selected layers (in ML density) and the average $\lambda$ (in our density) also increase. Additionally, our density provides a solution with a greater average $\lambda$ for all values of $\beta$.
Due to space constraints, we omit the results for the remaining datasets, which exhibit similar trends across all measures.

\begin{figure}[h]
    \centering
\includegraphics[width=0.65\linewidth]{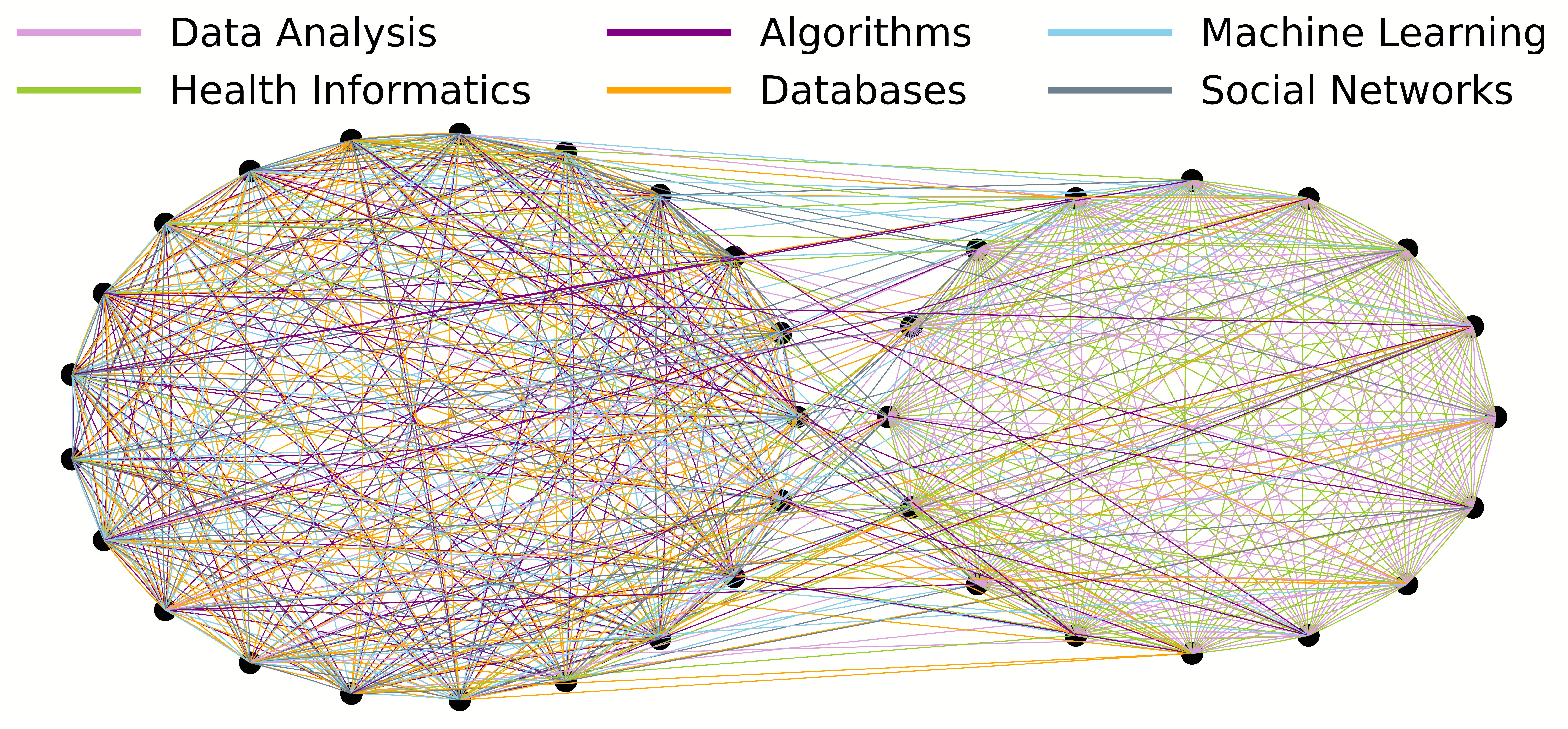}
    \caption{User
Multilayer densest subgraph based on our density in the DBLP dataset.}
    \label{fig:casestudy}
    \vspace{-2ex}
\end{figure}

\subsection{Case Study: DBLP}\label{app:DBLP}
As discussed in \autoref{sec:MDS}, existing density measures~\cite{MLcore, densest-common-subgraph} force all nodes to exhibit their high degree in all or a fixed subset of layers, which is a too hard constrainint and might cause missing some other dense structures. As the real-world evidence of this claim, in this part, we perform a case study on DBLP dataset~\cite{dblp}. In this dataset, each node is a researcher and two nodes are connected if they have published a paper with each other. The type of connections are obtained using LDA~\cite{LDA} algorithm on the topics and abstracts of the papers. Accordingly, each layer corresponds to collaboration network in a specific research topic. We report the found densest subgraph by our approach (Left) and \citet{MLcore} (Right) in Figure \ref{fig:casestudy}. While both approaches found the left community as a dense structure, the density measure proposed by \citet{MLcore} misses the other dense structure as they collaborated in different topics than the left community. Since this density metric forces all nodes to be active (having high degree) in a fixed set of layers, it misses this dense structure. 

\subsection{Implementation, Code, and Datasets} 
The source code, data, and/or other artifacts have been made available at \textcolor{c1}{\href{https://github.com/joint-em/FirmCore}{This link}}.

\end{document}